\documentclass[%
reprint,
superscriptaddress,
amsmath,amssymb,
aps,
]{revtex4-1}

\usepackage{graphicx}% Include figure files
\usepackage{dcolumn}% Align table columns on decimal point
\usepackage{bm}% bold math

\usepackage{csquotes}
\usepackage{graphicx}
\usepackage{float}%can set location option to [H] which is stricter than [h!]
\usepackage{hyperref}% skip to reference
\usepackage{textcomp}
\usepackage{colortbl}

\begin{document}
	
	\title[]{Towards temporal characterization of intense isolated attosecond pulses from relativistic surface high harmonics}
	
	\author{O. Jahn}
	\affiliation{Max-Planck-Institut f\"ur Quantenoptik, 85748 Garching, Germany}
	\affiliation{Department f\"ur Physik, Ludwig-Maximilians-Universit\"at M\"unchen, 85748 Garching, Germany}
	
	\author{V. E. Leshchenko}
	\email{vyacheslav.leshchenko@mpq.mpg.de}
	\affiliation{Max-Planck-Institut f\"ur Quantenoptik, 85748 Garching, Germany}
	\affiliation{Department f\"ur Physik, Ludwig-Maximilians-Universit\"at M\"unchen, 85748 Garching, Germany}
	
	\author{P. Tzallas}
	\affiliation{Foundation for Research and Technology-Hellas, Institute of Electronic Structure and Laser, Heraklion, Crete, Greece}
	\affiliation{ELI-ALPS, ELI-HU Non-Profit Ltd., Dugonics t\'er 13, Szeged 6720, Hungary} 
	
	\author{A. Kessel}
	\affiliation{Max-Planck-Institut f\"ur Quantenoptik, 85748 Garching, Germany}
	\affiliation{Department f\"ur Physik, Ludwig-Maximilians-Universit\"at M\"unchen, 85748 Garching, Germany}
	
	\author{M. Kr\"uger}
	\affiliation{Max-Planck-Institut f\"ur Quantenoptik, 85748 Garching, Germany}
	\affiliation{Department f\"ur Physik, Ludwig-Maximilians-Universit\"at M\"unchen, 85748 Garching, Germany}
	
	\author{A. M\"unzer}
	\affiliation{Max-Planck-Institut f\"ur Quantenoptik, 85748 Garching, Germany}
	\affiliation{Department f\"ur Physik, Ludwig-Maximilians-Universit\"at M\"unchen, 85748 Garching, Germany}
	
	\author{S. A. Trushin}
	\affiliation{Max-Planck-Institut f\"ur Quantenoptik, 85748 Garching, Germany}
	\affiliation{Department f\"ur Physik, Ludwig-Maximilians-Universit\"at M\"unchen, 85748 Garching, Germany}
	
	\author{M. Schultze}
	\affiliation{Max-Planck-Institut f\"ur Quantenoptik, 85748 Garching, Germany}
	
	\author{G. D. Tsakiris}
	\affiliation{Max-Planck-Institut f\"ur Quantenoptik, 85748 Garching, Germany}
	
	\author{S. Kahaly}
	\affiliation{ELI-ALPS, ELI-HU Non-Profit Ltd., Dugonics t\'er 13, Szeged 6720, Hungary}    
	
	\author{A. Guggenmos}
	\affiliation{Max-Planck-Institut f\"ur Quantenoptik, 85748 Garching, Germany}
	\affiliation{Department f\"ur Physik, Ludwig-Maximilians-Universit\"at M\"unchen, 85748 Garching, Germany}
	
	\author{D. Kormin}
	\affiliation{Max-Planck-Institut f\"ur Quantenoptik, 85748 Garching, Germany}
	\affiliation{Department f\"ur Physik, Ludwig-Maximilians-Universit\"at M\"unchen, 85748 Garching, Germany}
	
	\author{L. Veisz}
	\affiliation{Max-Planck-Institut f\"ur Quantenoptik, 85748 Garching, Germany}
	\affiliation{Department of Physics, Ume\r{a} University, Ume\r{a}, Sweden}
	
	\author{F. Krausz}
	\affiliation{Max-Planck-Institut f\"ur Quantenoptik, 85748 Garching, Germany}
	\affiliation{Department f\"ur Physik, Ludwig-Maximilians-Universit\"at M\"unchen, 85748 Garching, Germany}
	
	\author{Zs. Major}
	\altaffiliation{GSI Helmholtzzentrum f\"ur Schwerionenforschung GmbH, Planckstra{\ss}e 1, 64291 Darmstadt, Germany.\\Helmholtz-Institut Jena, Fr\"obelstieg 3, 07743 Jena, Germany.}
	\affiliation{Max-Planck-Institut f\"ur Quantenoptik, 85748 Garching, Germany}
	\affiliation{Department f\"ur Physik, Ludwig-Maximilians-Universit\"at M\"unchen, 85748 Garching, Germany}
	
	\author{S. Karsch}
	\email{stefan.karsch@mpq.mpg.de}
	\affiliation{Max-Planck-Institut f\"ur Quantenoptik, 85748 Garching, Germany}
	\affiliation{Department f\"ur Physik, Ludwig-Maximilians-Universit\"at M\"unchen, 85748 Garching, Germany}

%	\date{\today}
	
	\begin{abstract}

		Relativistic surface high harmonics have been considered a unique source for the generation of intense isolated attosecond pulses in the extreme ultra-violet (XUV) and X-ray spectral range. However, its experimental realization is still a challenging task requiring the identification
		of the optimum conditions for the generation of isolated attosecond pulses as well as their temporal characterization. Here, we demonstrate measurements in both directions. Particularly, we have made a first step towards the temporal characterization of the emitted XUV radiation by adapting the attosecond streak camera concept to identify the time domain characteristics of relativistic surface high harmonics. The results, supported by PIC simulations, set the upper limit for the averaged (over many shots) XUV duration to $<7$\,fs, even when driven by not CEP controlled relativistic few-cycle optical pulses. Moreover, by measuring the dependence of the spectrum of the relativistic surface high harmonics on the carrier envelope phase (CEP) of the driving infrared laser field, we experimentally determined the optimum conditions for the generation of intense isolated attosecond pulses.

		%\begin{description}
		
		
		%\end{description}
	\end{abstract}
	
	\maketitle

	The invention of sources of attosecond pulses based on high-order harmonic generation (HHG) \cite{PhysRevLett.70.774, hentschel2001, Paul2001} has opened the field of attosecond science \cite{Ferenc2009,Ferenc2014} with a wide range of potential applications \cite{REDUZZI2015}. Nowadays, attosecond science is mainly based on the HHG in gas media which allows the generation of isolated attosecond pulses on the nano- to few-micro Joule energy level with photon energies up to sub-keV. However, this approach has fundamental limitations determined by the ionization threshold of the gas medium \cite{Ferenc2009, Popmintchev2012}, leading to severe restrictions on the XUV flux especially at high photon energies.
	
	A way to overcome this limitation is to use relativistic harmonics generated by interaction of intense few-cycle laser fields with solid surfaces \cite{Plaja:98,George06,BGP,Heissler2012}. Theoretical predictions, based on the relativistic oscillating mirror (ROM) model \cite{George06}, have suggested that intense isolated attosecond pulses with up to few\thinspace keV photon energy can be generated when using few-cycle near-infrared (NIR) laser pulses with an intensity of $\sim 10^{20}\thinspace \text{W}/\text{cm}^2$. Therefore, ROM harmonics present one of the most promising attosecond sources for pump-probe studies in the X-ray spectral range. Yet, experimental obstacles associated mainly with the stringent requirements on the temporal contrast of the driving laser pulses have not yet allowed sufficient progress to realize the potential of this approach. However, recent progress in the development of laser systems based on optical parametric chirped pulse amplification (OPCPA) with pump pulse durations between 1 ps \cite{PFS_optica} and 80 ps \cite{Rivas2017} made the required pulse parameters available.
	Although the generation of isolated attosecond pulses from relativistic laser-plasma interactions driven by few-cycle optical pulses has been theoretically predicted using one-dimensional particle in cell (1D-PIC) simulations \cite{Heissler2012,Ma2015}, its experimental realization remains open.
	
	Here, using a 10\thinspace Hz repetition rate laser system \cite{PFS_optica}, delivering 7\thinspace fs pulses with 25\thinspace mJ energy at 900\thinspace nm central wavelength and better than $10^{-11}$ temporal contrast on the few-ps timescale, we experimentally investigate the dependence of the surface harmonics on the CEP of the driving field and demonstrate conditions under which isolated attosecond pulses can be generated. We compare our experimental data with 1D PIC simulations. The results are consistent with the outcome of the attosecond streaking measurements.
	
\begin{figure}[t]
		\includegraphics[width=\linewidth]{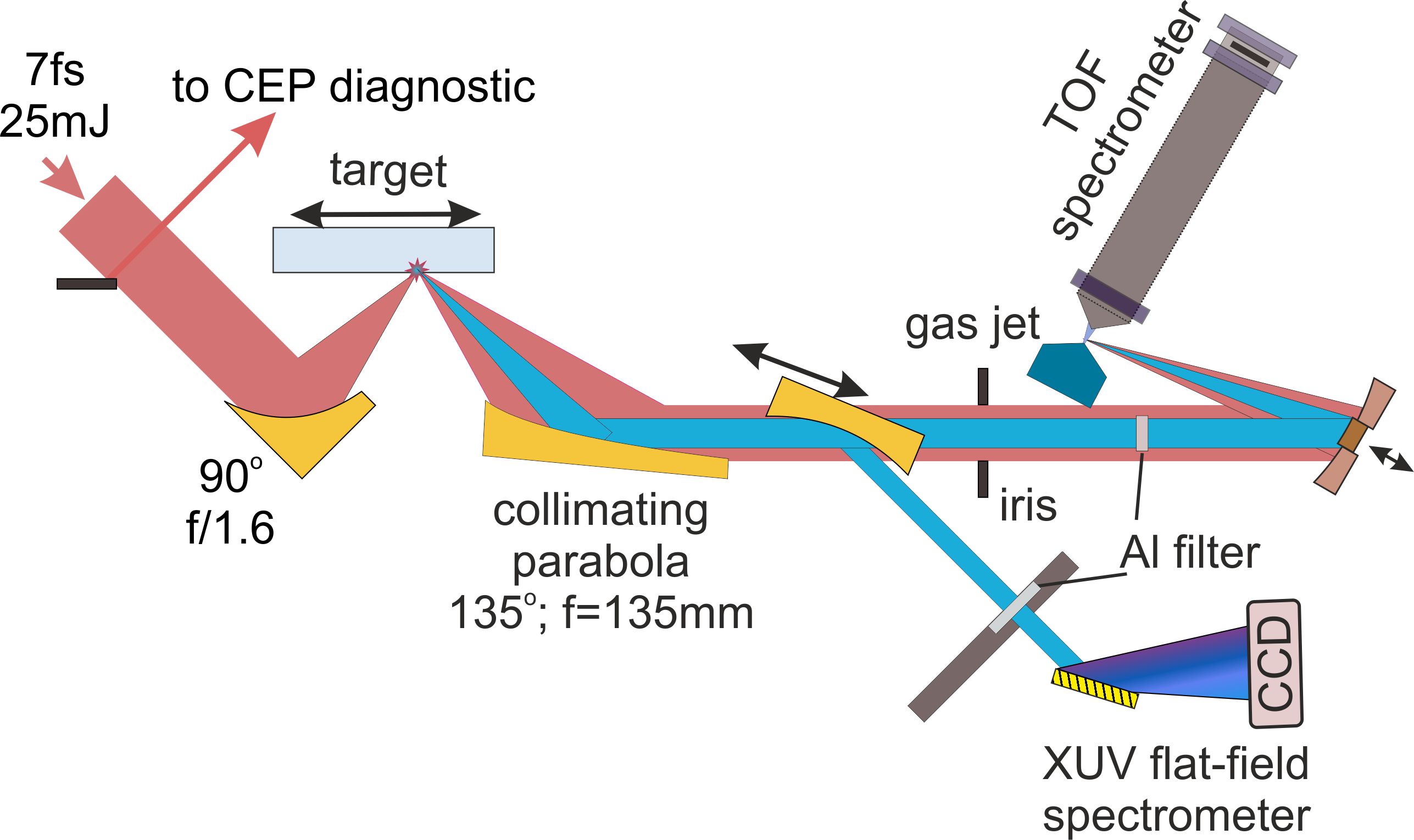}
		\centering
		\caption{Schematic setup of the SHHG experiment used for the measurement of the generated XUV spectrum and its temporal characterization by attosecond streaking. A two-component spherical mirror of 25 cm focal length was used for the streaking measurement. The inner part of the mirror with multilayer coating centered at 38\,eV with 3\,eV bandwidth and the outer part with silver coating were adjusted using piezo-actuators in order to optimize overlap and control the delay between the XUV and NIR pulses.}
		\label{fig:boat1}
	\end{figure}

	The experiment (Fig.~\ref{fig:boat1}) was performed by focusing p-polarized optical pulses onto a disk-like BK7 target under an incidence angle of 45\textdegree\ using a f/1.6 gold-coated 90\textdegree\ off-axis parabola. The peak intensity on target was $4\times10^{19}\thinspace\text{W}/\text{cm}^2$ resulting in a normalized vector potential of $a_0=4.8$, where $a_0^2=I\lambda ^2/(1.37\times 10^{18})$ with $I$ being the laser intensity in $\text{W}/\text{cm}^2$ and $\lambda$ the central wavelength in \textmu  \text{m}. The value of $a_0$, averaged over the full width at half maximum (FWHM) beam diameter, was around 3, therefore $a_0=3$ was used in our 1D PIC simulations. As it was shown in the previous experiments on relativistic surface HHG, it is most efficient when the pre-plasma has a scale length of $L_p\approx 0.1\lambda-0.2 \lambda$ \cite{Kahaly13,Jena2012,Dollar2013}. For this reason, a pre-pulse with an intensity of about $10^{15}\thinspace\text{W}/\text{cm}^2$ and an adjustable delay was introduced before the main pulse \cite{Kahaly13} to pre-ionize the target. The experimentally optimized pre-pulse delay was 3\thinspace ps which corresponds to $L_p\approx 0.2 \lambda$ \cite{plasma_expension2}. The spectrum of the generated harmonics was registered with a home-built single-shot XUV flat-field spectrometer after filtering out the residual fundamental radiation with a 200\thinspace nm thick aluminum filter. The maximum acquisition rate of the spectrometer XUV CCD camera limited the experimental repetition rate to 0.5\thinspace Hz. The CEP of the driving pulses was measured with a home-built single-shot f-2f spectral interferometer (Fig. 2(b) in \cite{CEP_tag}). The CEP diagnostic provides only relative values, therefore the absolute CEP was determined by fitting the experimental data to the simulations. In order to exclude the uncertainty introduced in the f-2f measurements \cite{f2f} by the 4\% energy instability of the driving pulses, the energy and the spectrum of the NIR pulses were recorded in parallel with the harmonic spectra and the f-2f signal (supplementary material \cite{sup_mat}). In this way, we were able to select the laser shots with energy instability $<$1\% and ensure that the CEP uncertainty is less than $\sim$200\,mrad \cite{f2f}.
	
\begin{figure}[t]
		\centering
		\includegraphics[width=\linewidth]{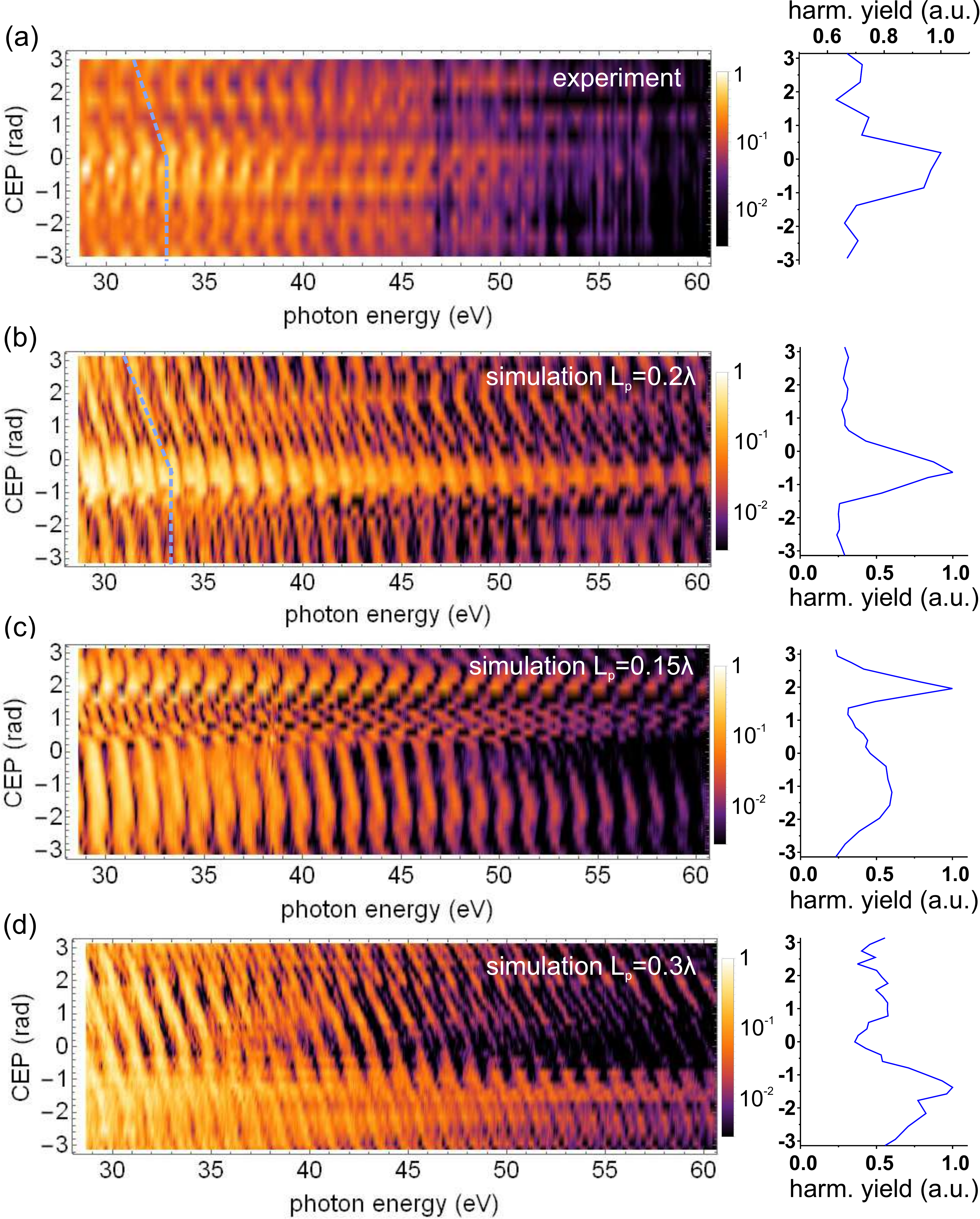}
		\caption{Dependence of the harmonic spectrum (left panels) and the integrated harmonic yield (right panels) on the CEP of the driving field. \textbf{(a)} Measurements (consisting of shots acquired over 20\thinspace min and sorted according to their CEP values). \textbf{(b--e)} PIC simulations for different plasma scale lengths $L_p$ (simulation parameters: $a_0=3$, $\tau=7\thinspace fs$, (b) $L_p=0.2\lambda$, (c) $L_p=0.15\lambda$, (d) $L_p=0.3\lambda$). CEP=0 corresponds to a cosine waveform with the maximum of the carrier wave in synchrony with the intensity envelope.}
		\label{fig:cep2d}
	\end{figure}

	The measured CEP dependence of the generated harmonics is shown in Fig.~\ref{fig:cep2d}(a). The corresponding result of a 1D PIC simulation \cite{LPIC} is presented in Fig.~\ref{fig:cep2d}(b). Both figures show the following features: I) a clear harmonic shift for positive CEP values (blue dashed lines) by about one harmonic order; II) the harmonic signal has a maximum ($\sim 4\thinspace \text{\textmu  J}$ measured in the presented 30--70\thinspace eV spectral range) at $\sim -0.5\thinspace \text{rad}$ and drops to a minimum when the phase changes by about $\pi/2$, i.e.\ at $\sim+1\thinspace \text{rad}$ and $\sim-2\thinspace \text{rad}$; III) around -2\thinspace rad, there is a subharmonic structure, namely small additional peaks between main harmonics. The energy was estimated by using the recorded harmonics spectra, the collection efficiency of the XUV relay optics, the grating diffraction efficiency, and the spectral detector response. Note that the harmonic spectra at $-\pi$ and $+\pi$ are the same, as they correspond to the same physical case. As expected \cite{Rivas2017,Borot2012}, the position of the harmonic peak (blue dashed line Fig.~\ref{fig:cep2d}(a-b)) moves to the next harmonic order in both experiment and simulation.
	
	For comparison, the simulation results for slightly different plasma scale lengths are presented in Fig.~\ref{fig:cep2d}(c-d). The CEP dependence of the harmonic position shift and of the integrated harmonic yield for these results obviously differ both from the experimental data and the simulations for $L_p=0.2\lambda$ that supports our estimation of the experimental plasma scale length of $L_p=0.2\lambda$. Therefore this approach can be used as a method to infer the plasma scale length from the CEP dependence of the relativistic surface high-order harmonics, which, although not direct and not single-shot as some other approaches \cite{plasma_expension2,Kahaly13}, can be useful for providing information on the plasma scale length in similar experiments without additional experimental effort.
	
	\begin{figure}[t]
		\centering
		\includegraphics[width=0.9\linewidth]{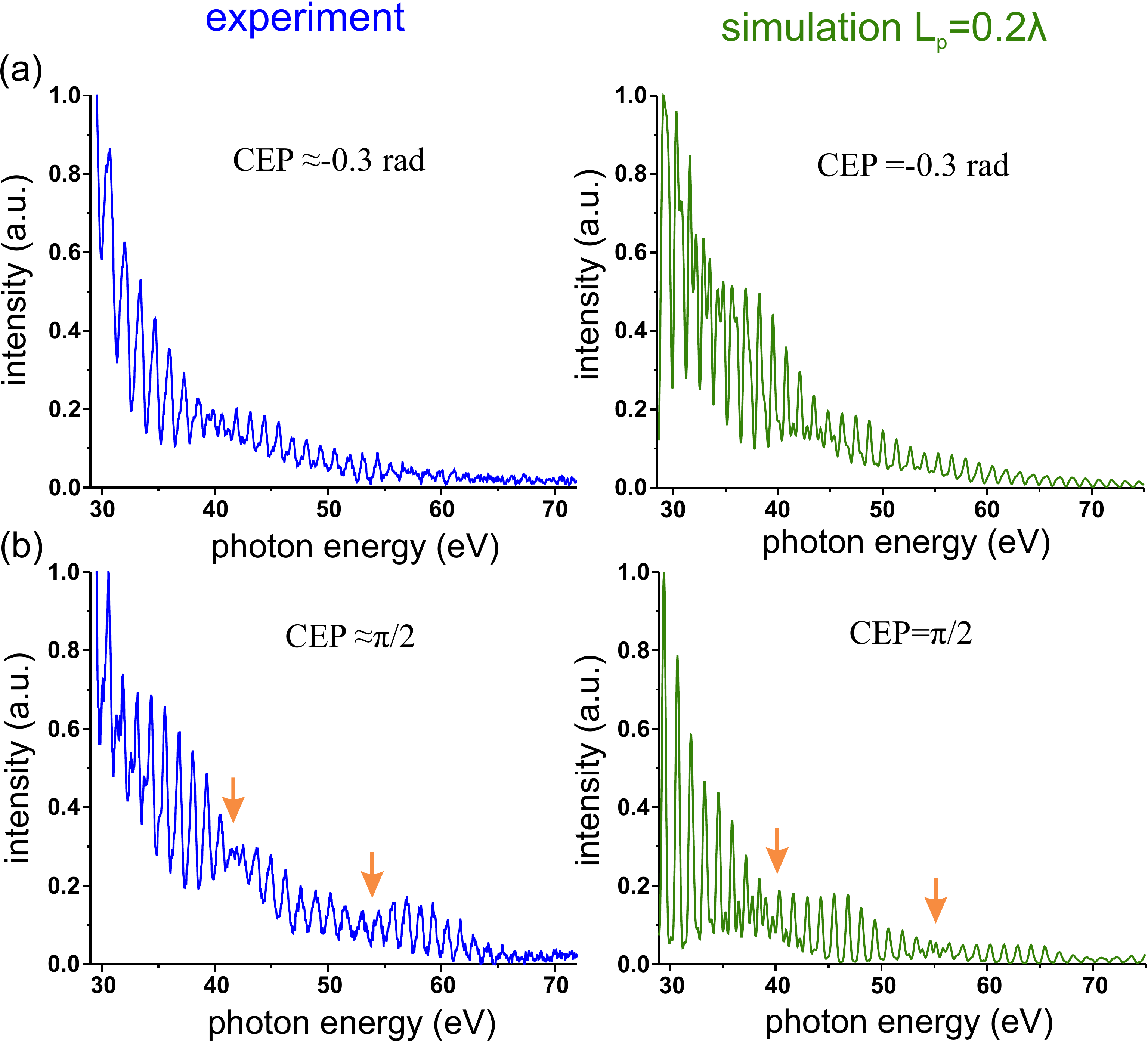}
		\caption{XUV spectra for different CEP values: single shot experimental results (left panels) and line-outs of simulation (right panels). \textbf{(a)} is the case with the smallest modulation depth and the best isolation degree (cf.\ Fig.~\ref{fig:cep_time}(d) and the main text for details). \textbf{(b)} shows the case with a pronounced beating structure (beating nodes are marked with arrows) which corresponds to the generation of three attosecond pulses with nearly equal amplitudes.}
		\label{fig:cep1d}
	\end{figure} 
	
	The agreement between experimental data and simulations for $L_p=0.2\lambda$ is even more prominently visible in the single-shot spectra shown in Fig.~\ref{fig:cep1d} where the slow modulation in the harmonic spectral amplitude and the modulation depth of the harmonic peaks are in fair agreement. 
	Note that harmonics above 33\,eV are generated by relativistic mechanisms \cite{BGP,CSE_Pukhov} because the CWE process \cite{CWE,Borot2012} can contribute only to the emission of photons with energy $<$33\,eV determined by the maximum plasma frequency when ionizing a BK7 target.	Furthermore, the measured spectra can be used for the evaluation of both the variation of the pulse spacing averaged over the train ($\Delta T$) as CEP changes and the value of the uneven spacing between pulses in the train ($\delta T$). The presence of the last effect is clear from the beating structure in the measured XUV spectra, which is most pronounced in the CEP=$\pi/2$ case (Fig.~\ref{fig:cep1d}(b)) and signifies that there are about three attosecond pulses with uneven temporal spacing \cite{Borot2012}. 
	
\begin{figure}[]
		
		\includegraphics[width=\linewidth]{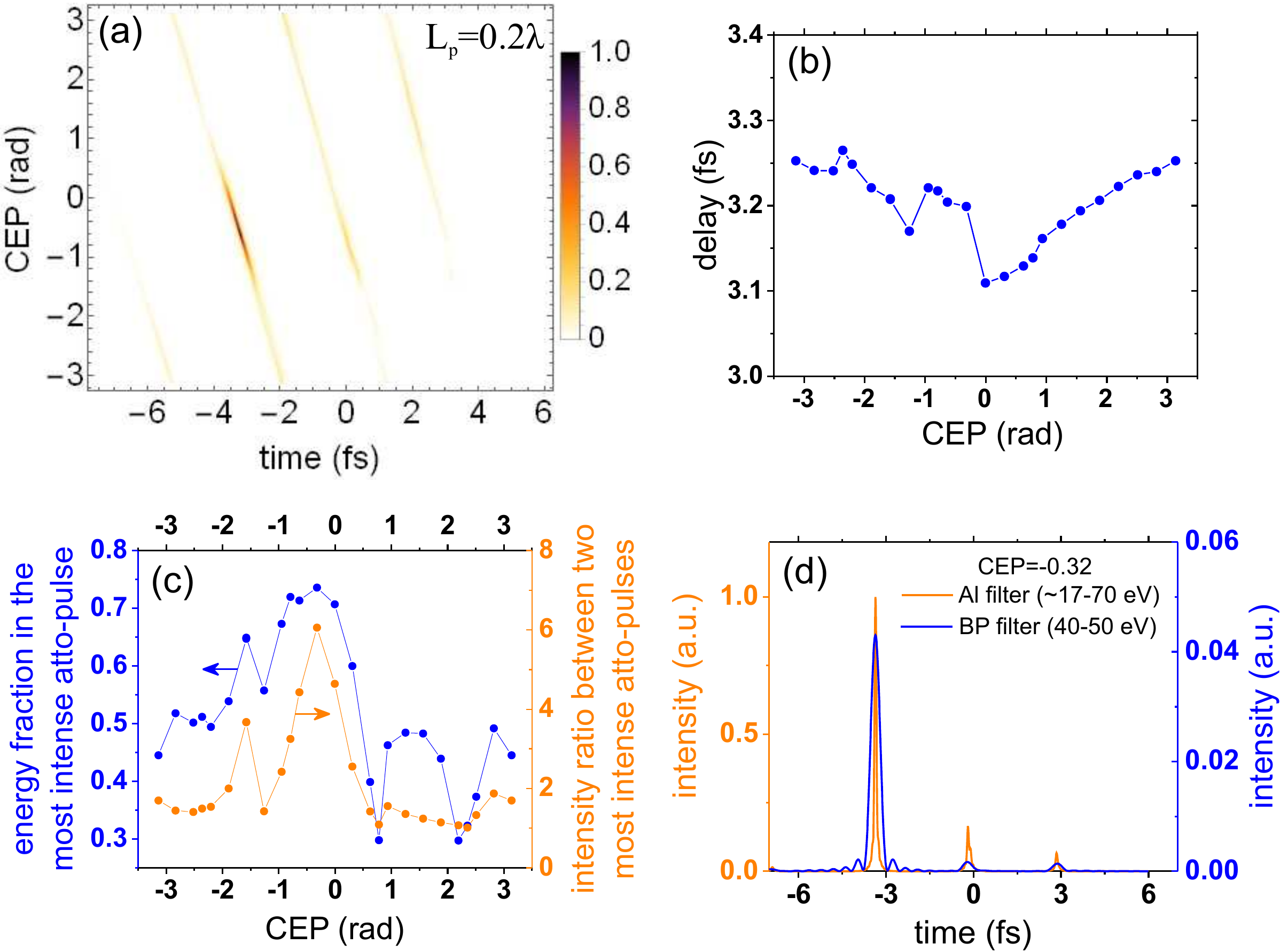}
		\caption{Temporal structure of the attosecond trace for $L_p=0.2\lambda$ from the 1D PIC simulations. \textbf{(a)} CEP dependence of the temporal structure of the attosecond pulse train (200 nm thick Al filter is applied). \textbf{(b)} Averaged delay ($\Delta T$) between attosecond pulses. \textbf{(c)} Energy (blue-dotted line) and intensity (orange-dotted line) ratio between the main attosecond pulse and the rest of the train. \textbf{(d)} Quasi-isolated attosecond pulse for CEP=-0.32\thinspace rad, 200\thinspace nm thick Al filter (orange line) and isolated attosecond pulse for the case of CEP=-0.32\thinspace rad and bandpass (BP) filter with 10\thinspace eV bandwidth centered at 45\thinspace eV (blue line). In these simulations $a_0=3$, $\tau=7\thinspace \text{fs}$, $L_p=0.2\lambda$.}
		\label{fig:cep_time}
	\end{figure}

	The measured CEP dependence (Fig.~\ref{fig:cep2d}(a)) shows that, for example, the n=25-th harmonic (35\thinspace eV) is shifted by one harmonic order, namely by 1.4\thinspace eV, when scanning the CEP from $-\pi$ to $\pi$. This corresponds to a change $\Delta T$ of the average pulse separation by $\Delta T=T_0/n=120\thinspace \text{as}$ (see supplementary material for details \cite{sup_mat}), where $T_0=3\thinspace fs$ is the period of the carrier. Nearly the same shift is observed in the simulations where the delay between attosecond pulses changes by 140\,as (Fig.~\ref{fig:cep_time}(b)).
	
	The period of the beating structure in the measured spectrum (Fig.~\ref{fig:cep1d}(b)) is about $f_{\text{beating}}=15$\thinspace eV which corresponds to $\delta T=1/f_{\text{beating}}=270$\thinspace as difference in the temporal spacing between attosecond pulses in the pulse train \cite{sup_mat}. The last result allows an estimation of the plasma denting \cite{denting}, namely the shift of the point of reflection from the plasma mirror by $\delta T\times c/(2\cos(45\textbf{\textdegree}))=57\thinspace \text{nm}$ \cite{sup_mat} during one optical cycle at the peak of the driving field under our experimental conditions. Thus a thorough analysis of the CEP harmonics spectra provides information on the plasma scale length and plasma dynamics during the interaction. The temporal structure of the emitted XUV radiation is discussed in the following.
	
	\begin{figure}[t]
		\includegraphics[width=\linewidth]{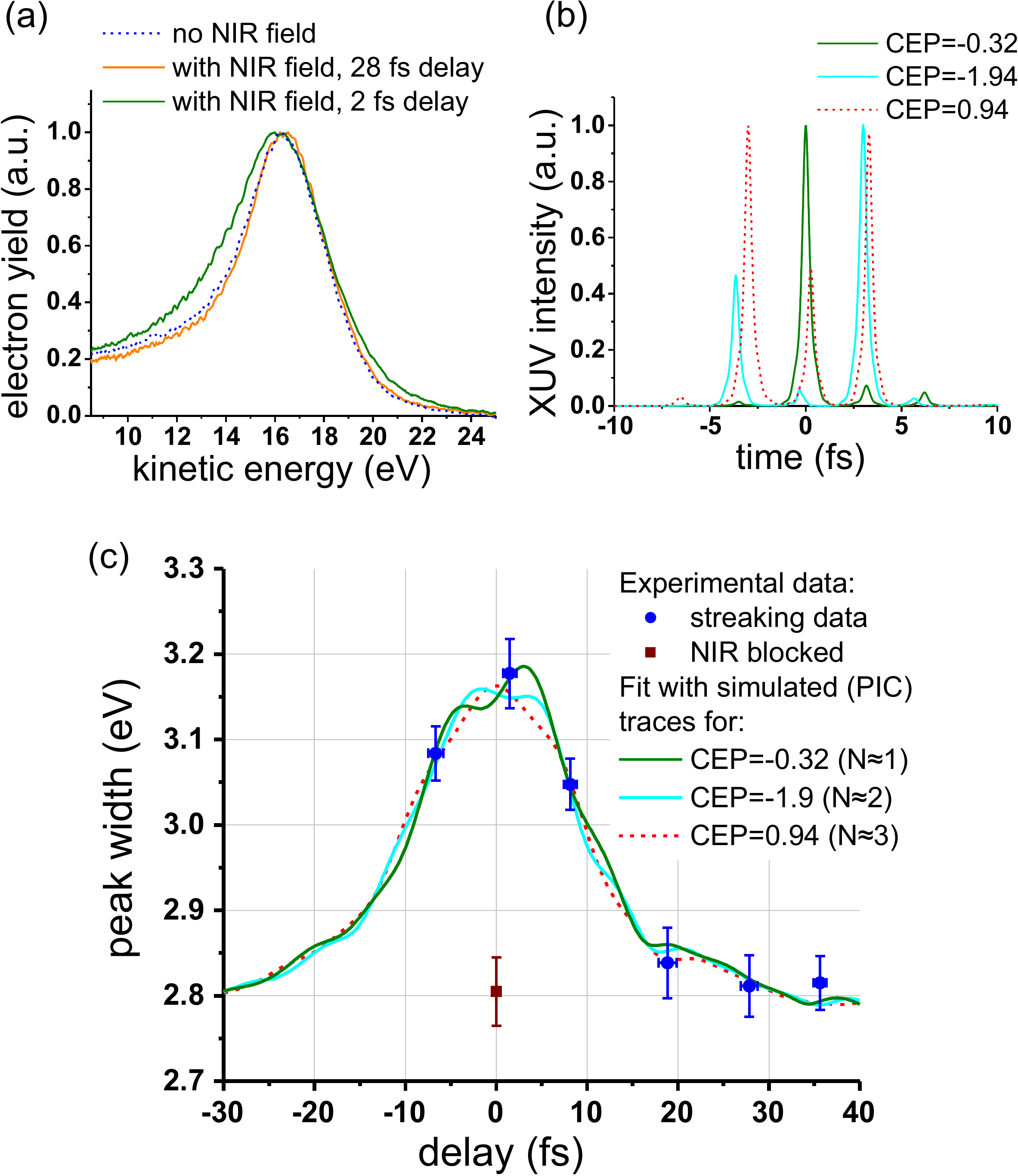}
		\caption{Streaking results. \textbf{(a)} Photoelectron (PE) signal recorded with (green and orange lines) and without (blue dashed line) streaking NIR field. The blue dashed line shows the PE signal produced by single photon ionization of Ne gas after the interaction with the focused XUV beam at the central photon energy of 38~eV. The green and orange lines show the PE signal when the XUV/NIR delay is 2\thinspace fs and 28\thinspace fs, respectively. \textbf{(b)} XUV radiation, used in the streaking simulations and obtained from the PIC simulations for $L_p=0.2\lambda$ after applying the transmission of the beamline determined by the Al filter and the XUV mirror. \textbf{(c)} Measured dependence of the PE peak width on the delay between XUV and NIR fields and simulated streaking traces using the attosecond trains from PIC simulations (see (b); one attosecond pulse (green solid line), two attosecond pulses (light blue solid line) and three attosecond pulses (red dashed line)). For each experimental point 100 shots were accumulated. Error bars depict one standard deviation from the mean value. The streaking simulations are averaged over results with random CEP.}
		\label{fig:streaking}
		
	\end{figure}
	
	Applying spectral transmission corresponding to a 200\thinspace nm thick Al filter, the temporal structure of the XUV radiation from the data set of Fig.~\ref{fig:cep2d}(b) is shown in Fig.~\ref{fig:cep_time}. From the energy ratio between the main attosecond pulse and the rest of the train, it is evident, that within nearly one half of the CEP range, namely between -2\thinspace rad and 0.5\thinspace rad, the XUV emission is mostly confined within one attosecond pulse. Using the intensity ratio between the main attosecond pulse and the rest of the train as the figure of merit for the degree of pulse isolation, the optimum CEP value under our experimental conditions is -0.3\thinspace rad. In this case, 74\% of the overall energy of the pulse train is contained within a quasi-isolated attosecond pulse (orange line in Fig.~\ref{fig:cep_time}(d)) which has a contrast of 0.16. The energy content and the contrast can be improved to 86\% and $4\times10^{-2}$, respectively, using a bandpass XUV filter with a 10\thinspace eV bandwidth centered at 45\thinspace eV that infers the generation of an isolated attosecond pulse (blue line in Fig.~\ref{fig:cep_time}(d)). Theoretical discussions of the mechanisms supporting the generation of isolated attosecond pulses even with few-cycle driving fields under conditions close to the ones in the performed experiments can be found in \cite{Ma2015,CSE_Pukhov,Gonoskov2018,Gonoskov2011} and in the supplementary material \cite{sup_mat}.
	
	While our experimental findings are in good agreement with the description provided by the simulations, a direct confirmation of the generation of an isolated attosecond pulse is the temporal characterization of the generated XUV radiation.	In our experiment, we have implemented the attosecond streaking approach \cite{Drescher2001,Itatani2002,Kienberger2004} (and supplementary material \cite{sup_mat}). The XUV and NIR beams were focused into a Ne gas jet by a two-component focusing spherical mirror (Fig.~\ref{fig:boat1}). The generated photoelectron (PE) spectrum was recorded by a time-of-flight (TOF) spectrometer. The duration of the NIR pulse on the gas jet was 14.5\,fs. Although the combination of a low repetition rate and a limited amount of shots (due to limited target size) did not allow us to record a full streaking trace, the measured data are sufficient to determine an upper limit of the XUV pulse duration. Typical recorded PE spectra are shown in Fig.~\ref{fig:streaking}(a). The streaking results exhibit a broadening of the PE distribution (Fig.~\ref{fig:streaking}(c)), with the FWHM width of the Gaussian fit of the trace being 19\thinspace fs. The experimental data are in agreement with the calculated streaking traces  \cite{Yakovlev05} using attosecond pulse trains consisting of up to 3 pulses (Fig.~\ref{fig:streaking}(b)), resulting in an overall XUV duration of $<7\thinspace \text{fs}$ that is consistent with the expected limit on the XUV train duration determined by the 7\,fs driving pulse. Additionally, streaking traces calculated for more than 3 XUV pulses in the train significantly deviate from the measured data (supplementary material \cite{sup_mat}). Finally we would like to note that the above measurement, to our knowledge, constitutes the first experimental demonstration of pump-probe studies using XUV radiation generated by relativistic surface high harmonics.
	
	In conclusion, by utilizing a high field few-cycle laser system, we have demonstrated the generation of intense relativistic surface high-order harmonics with few-fs averaged (over many shots) duration. The XUV pulse duration was deduced by streaking measurements and the results found to be in a fair agreement with 1D PIC simulations. The conditions supporting the generation of isolated attosecond XUV pulses are found by measuring the dependence of the harmonic spectrum on the CEP of the driving field. Also, the possibility to estimate the plasma dynamics during the laser-plasma interaction using the measured harmonics CEP dependence is demonstrated. The demonstrated XUV pulse energy level of several \textmu J in combination with the availability of CEP stable high intensity few-cycle laser pulses \cite{PFS_optica,Budriunas:17} commence a new era of experimental investigations in ultrafast non-linear XUV optics \cite{Ferenc2009,Tzallas2003} using relativistic surface high-order harmonics.
	
	%	\begin{acknowledgments}
	
	The authors acknowledge the contribution of former members of the team: I. Ahmad, S. Klingebiel, C. Wandt, A. Schwarz and C. Skrobol. Furthermore, we thank B. Bergues, E. Goulielmakis, M. Ossiander, F. Siegrist, U. Kleineberg, M. Weidman, T. T. Luu, Q. Liu, M. Kling, M. Gilljohann, V. Yakovlev and M. Ciappina for helpful discussions.
	This work was supported by DFG through the Cluster of Excellence \textquotedblleft Munich Center for Advanced Photonics\textquotedblright (MAP) (EXC 158) and TR-18 funding schemes; Euratom research and training program 2014-2018 under Grant agreement No. 633053 within the framework of the EUROfusion Consortium; \textquotedblleft International Max-Planck Research School of Advanced Photon Science\textquotedblright (IMPRS-APS), and the Max-Planck Society. S. Kahaly thanks ELI-ALPS supported by the European Union and co-financed by the European Regional Development Fund (ERDF) (GINOP-2.3.6-15-2015-00001).
	
	%	\end{acknowledgments}

%


\begin{thebibliography}{34}%
	\makeatletter
	\providecommand \@ifxundefined [1]{%
		\@ifx{#1\undefined}
	}%
	\providecommand \@ifnum [1]{%
		\ifnum #1\expandafter \@firstoftwo
		\else \expandafter \@secondoftwo
		\fi
	}%
	\providecommand \@ifx [1]{%
		\ifx #1\expandafter \@firstoftwo
		\else \expandafter \@secondoftwo
		\fi
	}%
	\providecommand \natexlab [1]{#1}%
	\providecommand \enquote  [1]{``#1''}%
	\providecommand \bibnamefont  [1]{#1}%
	\providecommand \bibfnamefont [1]{#1}%
	\providecommand \citenamefont [1]{#1}%
	\providecommand \href@noop [0]{\@secondoftwo}%
	\providecommand \href [0]{\begingroup \@sanitize@url \@href}%
	\providecommand \@href[1]{\@@startlink{#1}\@@href}%
	\providecommand \@@href[1]{\endgroup#1\@@endlink}%
	\providecommand \@sanitize@url [0]{\catcode `\\12\catcode `\$12\catcode
		`\&12\catcode `\#12\catcode `\^12\catcode `\_12\catcode `\%12\relax}%
	\providecommand \@@startlink[1]{}%
	\providecommand \@@endlink[0]{}%
	\providecommand \url  [0]{\begingroup\@sanitize@url \@url }%
	\providecommand \@url [1]{\endgroup\@href {#1}{\urlprefix }}%
	\providecommand \urlprefix  [0]{URL }%
	\providecommand \Eprint [0]{\href }%
	\providecommand \doibase [0]{http://dx.doi.org/}%
	\providecommand \selectlanguage [0]{\@gobble}%
	\providecommand \bibinfo  [0]{\@secondoftwo}%
	\providecommand \bibfield  [0]{\@secondoftwo}%
	\providecommand \translation [1]{[#1]}%
	\providecommand \BibitemOpen [0]{}%
	\providecommand \bibitemStop [0]{}%
	\providecommand \bibitemNoStop [0]{.\EOS\space}%
	\providecommand \EOS [0]{\spacefactor3000\relax}%
	\providecommand \BibitemShut  [1]{\csname bibitem#1\endcsname}%
	\let\auto@bib@innerbib\@empty
	%</preamble>
	\bibitem [{\citenamefont {L'Huillier}\ and\ \citenamefont
		{Balcou}(1993)}]{PhysRevLett.70.774}%
	\BibitemOpen
	\bibfield  {author} {\bibinfo {author} {\bibfnamefont {A.}~\bibnamefont
			{L'Huillier}}\ and\ \bibinfo {author} {\bibfnamefont {P.}~\bibnamefont
			{Balcou}},\ }\href {\doibase 10.1103/PhysRevLett.70.774} {\bibfield
		{journal} {\bibinfo  {journal} {Phys. Rev. Lett.}\ }\textbf {\bibinfo
			{volume} {70}},\ \bibinfo {pages} {774} (\bibinfo {year} {1993})}\BibitemShut
	{NoStop}%
	\bibitem [{\citenamefont {Hentschel}\ \emph {et~al.}(2001)\citenamefont
		{Hentschel}, \citenamefont {Kienberger}, \citenamefont {Spielmann},
		\citenamefont {Reider}, \citenamefont {Milosevic}, \citenamefont {Brabec},
		\citenamefont {Corkum}, \citenamefont {Heinzmann}, \citenamefont {Drescher},\
		and\ \citenamefont {Krausz}}]{hentschel2001}%
	\BibitemOpen
	\bibfield  {author} {\bibinfo {author} {\bibfnamefont {M.}~\bibnamefont
			{Hentschel}}, \bibinfo {author} {\bibfnamefont {R.}~\bibnamefont
			{Kienberger}}, \bibinfo {author} {\bibfnamefont {C.}~\bibnamefont
			{Spielmann}}, \bibinfo {author} {\bibfnamefont {G.~A.}\ \bibnamefont
			{Reider}}, \bibinfo {author} {\bibfnamefont {N.}~\bibnamefont {Milosevic}},
		\bibinfo {author} {\bibfnamefont {T.}~\bibnamefont {Brabec}}, \bibinfo
		{author} {\bibfnamefont {P.}~\bibnamefont {Corkum}}, \bibinfo {author}
		{\bibfnamefont {U.}~\bibnamefont {Heinzmann}}, \bibinfo {author}
		{\bibfnamefont {M.}~\bibnamefont {Drescher}}, \ and\ \bibinfo {author}
		{\bibfnamefont {F.}~\bibnamefont {Krausz}},\ }\href@noop {} {\bibfield
		{journal} {\bibinfo  {journal} {Nature}\ }\textbf {\bibinfo {volume} {414}},\
		\bibinfo {pages} {509} (\bibinfo {year} {2001})}\BibitemShut {NoStop}%
	\bibitem [{\citenamefont {Paul}\ \emph {et~al.}(2001)\citenamefont {Paul},
		\citenamefont {Toma}, \citenamefont {Breger}, \citenamefont {Mullot},
		\citenamefont {Aug{\'e}}, \citenamefont {Balcou}, \citenamefont {Muller},\
		and\ \citenamefont {Agostini}}]{Paul2001}%
	\BibitemOpen
	\bibfield  {author} {\bibinfo {author} {\bibfnamefont {P.~M.}\ \bibnamefont
			{Paul}}, \bibinfo {author} {\bibfnamefont {E.~S.}\ \bibnamefont {Toma}},
		\bibinfo {author} {\bibfnamefont {P.}~\bibnamefont {Breger}}, \bibinfo
		{author} {\bibfnamefont {G.}~\bibnamefont {Mullot}}, \bibinfo {author}
		{\bibfnamefont {F.}~\bibnamefont {Aug{\'e}}}, \bibinfo {author}
		{\bibfnamefont {P.}~\bibnamefont {Balcou}}, \bibinfo {author} {\bibfnamefont
			{H.~G.}\ \bibnamefont {Muller}}, \ and\ \bibinfo {author} {\bibfnamefont
			{P.}~\bibnamefont {Agostini}},\ }\href {\doibase 10.1126/science.1059413}
	{\bibfield  {journal} {\bibinfo  {journal} {Science}\ }\textbf {\bibinfo
			{volume} {292}},\ \bibinfo {pages} {1689} (\bibinfo {year}
		{2001})}\BibitemShut {NoStop}%
	\bibitem [{\citenamefont {Krausz}\ and\ \citenamefont
		{Ivanov}(2009)}]{Ferenc2009}%
	\BibitemOpen
	\bibfield  {author} {\bibinfo {author} {\bibfnamefont {F.}~\bibnamefont
			{Krausz}}\ and\ \bibinfo {author} {\bibfnamefont {M.}~\bibnamefont
			{Ivanov}},\ }\href {\doibase 10.1103/RevModPhys.81.163} {\bibfield  {journal}
		{\bibinfo  {journal} {Rev. Mod. Phys.}\ }\textbf {\bibinfo {volume} {81}},\
		\bibinfo {pages} {163} (\bibinfo {year} {2009})}\BibitemShut {NoStop}%
	\bibitem [{\citenamefont {Krausz}\ and\ \citenamefont
		{Stockman}(2014)}]{Ferenc2014}%
	\BibitemOpen
	\bibfield  {author} {\bibinfo {author} {\bibfnamefont {F.}~\bibnamefont
			{Krausz}}\ and\ \bibinfo {author} {\bibfnamefont {M.~I.}\ \bibnamefont
			{Stockman}},\ }\href@noop {} {\bibfield  {journal} {\bibinfo  {journal}
			{Nature Photonics}\ }\textbf {\bibinfo {volume} {8}},\ \bibinfo {pages} {205}
		(\bibinfo {year} {2014})}\BibitemShut {NoStop}%
	\bibitem [{\citenamefont {Reduzzi}\ \emph {et~al.}(2015)\citenamefont
		{Reduzzi}, \citenamefont {Carpeggiani}, \citenamefont {Kühn}, \citenamefont
		{Calegari}, \citenamefont {Nisoli}, \citenamefont {Stagira}, \citenamefont
		{Vozzi}, \citenamefont {Dombi}, \citenamefont {Kahaly}, \citenamefont
		{Tzallas}, \citenamefont {Charalambidis}, \citenamefont {Varju},
		\citenamefont {Osvay},\ and\ \citenamefont {Sansone}}]{REDUZZI2015}%
	\BibitemOpen
	\bibfield  {author} {\bibinfo {author} {\bibfnamefont {M.}~\bibnamefont
			{Reduzzi}}, \bibinfo {author} {\bibfnamefont {P.}~\bibnamefont
			{Carpeggiani}}, \bibinfo {author} {\bibfnamefont {S.}~\bibnamefont {Kühn}},
		\bibinfo {author} {\bibfnamefont {F.}~\bibnamefont {Calegari}}, \bibinfo
		{author} {\bibfnamefont {M.}~\bibnamefont {Nisoli}}, \bibinfo {author}
		{\bibfnamefont {S.}~\bibnamefont {Stagira}}, \bibinfo {author} {\bibfnamefont
			{C.}~\bibnamefont {Vozzi}}, \bibinfo {author} {\bibfnamefont
			{P.}~\bibnamefont {Dombi}}, \bibinfo {author} {\bibfnamefont
			{S.}~\bibnamefont {Kahaly}}, \bibinfo {author} {\bibfnamefont
			{P.}~\bibnamefont {Tzallas}}, \bibinfo {author} {\bibfnamefont
			{D.}~\bibnamefont {Charalambidis}}, \bibinfo {author} {\bibfnamefont
			{K.}~\bibnamefont {Varju}}, \bibinfo {author} {\bibfnamefont
			{K.}~\bibnamefont {Osvay}}, \ and\ \bibinfo {author} {\bibfnamefont
			{G.}~\bibnamefont {Sansone}},\ }\href {\doibase
		https://doi.org/10.1016/j.elspec.2015.09.002} {\bibfield  {journal} {\bibinfo
			{journal} {Journal of Electron Spectroscopy and Related Phenomena}\ }\textbf
		{\bibinfo {volume} {204}},\ \bibinfo {pages} {257} (\bibinfo {year}
		{2015})}\BibitemShut {NoStop}%
	\bibitem [{\citenamefont {Popmintchev}\ \emph {et~al.}(2012)\citenamefont
		{Popmintchev}, \citenamefont {Chen}, \citenamefont {Popmintchev},
		\citenamefont {Arpin}, \citenamefont {Brown}, \citenamefont {Ali{\v
				s}auskas}, \citenamefont {Andriukaitis}, \citenamefont {Bal{\v c}iunas},
		\citenamefont {M{\"u}cke}, \citenamefont {Pugzlys}, \citenamefont {Baltu{\v
				s}ka}, \citenamefont {Shim}, \citenamefont {Schrauth}, \citenamefont {Gaeta},
		\citenamefont {Hern{\'a}ndez-Garc{\'\i}a}, \citenamefont {Plaja},
		\citenamefont {Becker}, \citenamefont {Jaron-Becker}, \citenamefont
		{Murnane},\ and\ \citenamefont {Kapteyn}}]{Popmintchev2012}%
	\BibitemOpen
	\bibfield  {author} {\bibinfo {author} {\bibfnamefont {T.}~\bibnamefont
			{Popmintchev}}, \bibinfo {author} {\bibfnamefont {M.-C.}\ \bibnamefont
			{Chen}}, \bibinfo {author} {\bibfnamefont {D.}~\bibnamefont {Popmintchev}},
		\bibinfo {author} {\bibfnamefont {P.}~\bibnamefont {Arpin}}, \bibinfo
		{author} {\bibfnamefont {S.}~\bibnamefont {Brown}}, \bibinfo {author}
		{\bibfnamefont {S.}~\bibnamefont {Ali{\v s}auskas}}, \bibinfo {author}
		{\bibfnamefont {G.}~\bibnamefont {Andriukaitis}}, \bibinfo {author}
		{\bibfnamefont {T.}~\bibnamefont {Bal{\v c}iunas}}, \bibinfo {author}
		{\bibfnamefont {O.~D.}\ \bibnamefont {M{\"u}cke}}, \bibinfo {author}
		{\bibfnamefont {A.}~\bibnamefont {Pugzlys}}, \bibinfo {author} {\bibfnamefont
			{A.}~\bibnamefont {Baltu{\v s}ka}}, \bibinfo {author} {\bibfnamefont
			{B.}~\bibnamefont {Shim}}, \bibinfo {author} {\bibfnamefont {S.~E.}\
			\bibnamefont {Schrauth}}, \bibinfo {author} {\bibfnamefont {A.}~\bibnamefont
			{Gaeta}}, \bibinfo {author} {\bibfnamefont {C.}~\bibnamefont
			{Hern{\'a}ndez-Garc{\'\i}a}}, \bibinfo {author} {\bibfnamefont
			{L.}~\bibnamefont {Plaja}}, \bibinfo {author} {\bibfnamefont
			{A.}~\bibnamefont {Becker}}, \bibinfo {author} {\bibfnamefont
			{A.}~\bibnamefont {Jaron-Becker}}, \bibinfo {author} {\bibfnamefont {M.~M.}\
			\bibnamefont {Murnane}}, \ and\ \bibinfo {author} {\bibfnamefont {H.~C.}\
			\bibnamefont {Kapteyn}},\ }\href {\doibase 10.1126/science.1218497}
	{\bibfield  {journal} {\bibinfo  {journal} {Science}\ }\textbf {\bibinfo
			{volume} {336}},\ \bibinfo {pages} {1287} (\bibinfo {year}
		{2012})}\BibitemShut {NoStop}%
	\bibitem [{\citenamefont {Plaja}\ \emph {et~al.}(1998)\citenamefont {Plaja},
		\citenamefont {Roso}, \citenamefont {Rza̧\.{z}ewski},\ and\ \citenamefont
		{Lewenstein}}]{Plaja:98}%
	\BibitemOpen
	\bibfield  {author} {\bibinfo {author} {\bibfnamefont {L.}~\bibnamefont
			{Plaja}}, \bibinfo {author} {\bibfnamefont {L.}~\bibnamefont {Roso}},
		\bibinfo {author} {\bibfnamefont {K.}~\bibnamefont {Rza̧\.{z}ewski}}, \ and\
		\bibinfo {author} {\bibfnamefont {M.}~\bibnamefont {Lewenstein}},\ }\href
	{\doibase 10.1364/JOSAB.15.001904} {\bibfield  {journal} {\bibinfo  {journal}
			{J. Opt. Soc. Am. B}\ }\textbf {\bibinfo {volume} {15}},\ \bibinfo {pages}
		{1904} (\bibinfo {year} {1998})}\BibitemShut {NoStop}%
	\bibitem [{\citenamefont {Tsakiris}\ \emph {et~al.}(2006)\citenamefont
		{Tsakiris}, \citenamefont {Eidmann}, \citenamefont {ter Vehn},\ and\
		\citenamefont {Krausz}}]{George06}%
	\BibitemOpen
	\bibfield  {author} {\bibinfo {author} {\bibfnamefont {G.~D.}\ \bibnamefont
			{Tsakiris}}, \bibinfo {author} {\bibfnamefont {K.}~\bibnamefont {Eidmann}},
		\bibinfo {author} {\bibfnamefont {J.~M.}\ \bibnamefont {ter Vehn}}, \ and\
		\bibinfo {author} {\bibfnamefont {F.}~\bibnamefont {Krausz}},\ }\href
	{http://stacks.iop.org/1367-2630/8/i=1/a=019} {\bibfield  {journal} {\bibinfo
			{journal} {New Journal of Physics}\ }\textbf {\bibinfo {volume} {8}},\
		\bibinfo {pages} {19} (\bibinfo {year} {2006})}\BibitemShut {NoStop}%
	\bibitem [{\citenamefont {Baeva}\ \emph {et~al.}(2006)\citenamefont {Baeva},
		\citenamefont {Gordienko},\ and\ \citenamefont {Pukhov}}]{BGP}%
	\BibitemOpen
	\bibfield  {author} {\bibinfo {author} {\bibfnamefont {T.}~\bibnamefont
			{Baeva}}, \bibinfo {author} {\bibfnamefont {S.}~\bibnamefont {Gordienko}}, \
		and\ \bibinfo {author} {\bibfnamefont {A.}~\bibnamefont {Pukhov}},\
	}\href@noop {} {\bibfield  {journal} {\bibinfo  {journal} {Phys. Rev. E}\
		}\textbf {\bibinfo {volume} {74}},\ \bibinfo {pages} {046404} (\bibinfo
		{year} {2006})}\BibitemShut {NoStop}%
	\bibitem [{\citenamefont {Heissler}\ \emph {et~al.}(2012)\citenamefont
		{Heissler}, \citenamefont {H\"orlein}, \citenamefont {Mikhailova},
		\citenamefont {Waldecker}, \citenamefont {Tzallas}, \citenamefont {Buck},
		\citenamefont {Schmid}, \citenamefont {Sears}, \citenamefont {Krausz},
		\citenamefont {Veisz}, \citenamefont {Zepf},\ and\ \citenamefont
		{Tsakiris}}]{Heissler2012}%
	\BibitemOpen
	\bibfield  {author} {\bibinfo {author} {\bibfnamefont {P.}~\bibnamefont
			{Heissler}}, \bibinfo {author} {\bibfnamefont {R.}~\bibnamefont {H\"orlein}},
		\bibinfo {author} {\bibfnamefont {J.~M.}\ \bibnamefont {Mikhailova}},
		\bibinfo {author} {\bibfnamefont {L.}~\bibnamefont {Waldecker}}, \bibinfo
		{author} {\bibfnamefont {P.}~\bibnamefont {Tzallas}}, \bibinfo {author}
		{\bibfnamefont {A.}~\bibnamefont {Buck}}, \bibinfo {author} {\bibfnamefont
			{K.}~\bibnamefont {Schmid}}, \bibinfo {author} {\bibfnamefont {C.~M.~S.}\
			\bibnamefont {Sears}}, \bibinfo {author} {\bibfnamefont {F.}~\bibnamefont
			{Krausz}}, \bibinfo {author} {\bibfnamefont {L.}~\bibnamefont {Veisz}},
		\bibinfo {author} {\bibfnamefont {M.}~\bibnamefont {Zepf}}, \ and\ \bibinfo
		{author} {\bibfnamefont {G.~D.}\ \bibnamefont {Tsakiris}},\ }\href {\doibase
		10.1103/PhysRevLett.108.235003} {\bibfield  {journal} {\bibinfo  {journal}
			{Phys. Rev. Lett.}\ }\textbf {\bibinfo {volume} {108}},\ \bibinfo {pages}
		{235003} (\bibinfo {year} {2012})}\BibitemShut {NoStop}%
	\bibitem [{\citenamefont {Kessel}\ \emph {et~al.}(2018)\citenamefont {Kessel},
		\citenamefont {Leshchenko}, \citenamefont {Jahn}, \citenamefont {Kr\"{u}ger},
		\citenamefont {M\"{u}nzer}, \citenamefont {Schwarz}, \citenamefont {Pervak},
		\citenamefont {Trubetskov}, \citenamefont {Trushin}, \citenamefont {Krausz},
		\citenamefont {Major},\ and\ \citenamefont {Karsch}}]{PFS_optica}%
	\BibitemOpen
	\bibfield  {author} {\bibinfo {author} {\bibfnamefont {A.}~\bibnamefont
			{Kessel}}, \bibinfo {author} {\bibfnamefont {V.~E.}\ \bibnamefont
			{Leshchenko}}, \bibinfo {author} {\bibfnamefont {O.}~\bibnamefont {Jahn}},
		\bibinfo {author} {\bibfnamefont {M.}~\bibnamefont {Kr\"{u}ger}}, \bibinfo
		{author} {\bibfnamefont {A.}~\bibnamefont {M\"{u}nzer}}, \bibinfo {author}
		{\bibfnamefont {A.}~\bibnamefont {Schwarz}}, \bibinfo {author} {\bibfnamefont
			{V.}~\bibnamefont {Pervak}}, \bibinfo {author} {\bibfnamefont
			{M.}~\bibnamefont {Trubetskov}}, \bibinfo {author} {\bibfnamefont {S.~A.}\
			\bibnamefont {Trushin}}, \bibinfo {author} {\bibfnamefont {F.}~\bibnamefont
			{Krausz}}, \bibinfo {author} {\bibfnamefont {Z.}~\bibnamefont {Major}}, \
		and\ \bibinfo {author} {\bibfnamefont {S.}~\bibnamefont {Karsch}},\
	}\href@noop {} {\bibfield  {journal} {\bibinfo  {journal} {Optica}\ }\textbf
		{\bibinfo {volume} {5}},\ \bibinfo {pages} {434} (\bibinfo {year}
		{2018})}\BibitemShut {NoStop}%
	\bibitem [{\citenamefont {Rivas}\ \emph {et~al.}(2017)\citenamefont {Rivas},
		\citenamefont {Borot}, \citenamefont {Cardenas}, \citenamefont {Marcus},
		\citenamefont {Gu}, \citenamefont {Herrmann}, \citenamefont {Xu},
		\citenamefont {Tan}, \citenamefont {Kormin}, \citenamefont {Ma},
		\citenamefont {Dallari}, \citenamefont {Tsakiris}, \citenamefont
		{F{\"o}ldes}, \citenamefont {Chou}, \citenamefont {Weidman}, \citenamefont
		{Bergues}, \citenamefont {Wittmann}, \citenamefont {Schr{\"o}der},
		\citenamefont {Tzallas}, \citenamefont {Charalambidis}, \citenamefont
		{Razskazovskaya}, \citenamefont {Pervak}, \citenamefont {Krausz},\ and\
		\citenamefont {Veisz}}]{Rivas2017}%
	\BibitemOpen
	\bibfield  {author} {\bibinfo {author} {\bibfnamefont {D.~E.}\ \bibnamefont
			{Rivas}}, \bibinfo {author} {\bibfnamefont {A.}~\bibnamefont {Borot}},
		\bibinfo {author} {\bibfnamefont {D.~E.}\ \bibnamefont {Cardenas}}, \bibinfo
		{author} {\bibfnamefont {G.}~\bibnamefont {Marcus}}, \bibinfo {author}
		{\bibfnamefont {X.}~\bibnamefont {Gu}}, \bibinfo {author} {\bibfnamefont
			{D.}~\bibnamefont {Herrmann}}, \bibinfo {author} {\bibfnamefont
			{J.}~\bibnamefont {Xu}}, \bibinfo {author} {\bibfnamefont {J.}~\bibnamefont
			{Tan}}, \bibinfo {author} {\bibfnamefont {D.}~\bibnamefont {Kormin}},
		\bibinfo {author} {\bibfnamefont {G.}~\bibnamefont {Ma}}, \bibinfo {author}
		{\bibfnamefont {W.}~\bibnamefont {Dallari}}, \bibinfo {author} {\bibfnamefont
			{G.~D.}\ \bibnamefont {Tsakiris}}, \bibinfo {author} {\bibfnamefont {I.~B.}\
			\bibnamefont {F{\"o}ldes}}, \bibinfo {author} {\bibfnamefont {S.~w.}\
			\bibnamefont {Chou}}, \bibinfo {author} {\bibfnamefont {M.}~\bibnamefont
			{Weidman}}, \bibinfo {author} {\bibfnamefont {B.}~\bibnamefont {Bergues}},
		\bibinfo {author} {\bibfnamefont {T.}~\bibnamefont {Wittmann}}, \bibinfo
		{author} {\bibfnamefont {H.}~\bibnamefont {Schr{\"o}der}}, \bibinfo {author}
		{\bibfnamefont {P.}~\bibnamefont {Tzallas}}, \bibinfo {author} {\bibfnamefont
			{D.}~\bibnamefont {Charalambidis}}, \bibinfo {author} {\bibfnamefont
			{O.}~\bibnamefont {Razskazovskaya}}, \bibinfo {author} {\bibfnamefont
			{V.}~\bibnamefont {Pervak}}, \bibinfo {author} {\bibfnamefont
			{F.}~\bibnamefont {Krausz}}, \ and\ \bibinfo {author} {\bibfnamefont
			{L.}~\bibnamefont {Veisz}},\ }\href {\doibase 10.1038/s41598-017-05082-w}
	{\bibfield  {journal} {\bibinfo  {journal} {Scientific Reports}\ }\textbf
		{\bibinfo {volume} {7}},\ \bibinfo {pages} {5224} (\bibinfo {year}
		{2017})}\BibitemShut {NoStop}%
	\bibitem [{\citenamefont {Ma}\ \emph {et~al.}(2015)\citenamefont {Ma},
		\citenamefont {Dallari}, \citenamefont {Borot}, \citenamefont {Krausz},
		\citenamefont {Yu}, \citenamefont {Tsakiris},\ and\ \citenamefont
		{Veisz}}]{Ma2015}%
	\BibitemOpen
	\bibfield  {author} {\bibinfo {author} {\bibfnamefont {G.}~\bibnamefont
			{Ma}}, \bibinfo {author} {\bibfnamefont {W.}~\bibnamefont {Dallari}},
		\bibinfo {author} {\bibfnamefont {A.}~\bibnamefont {Borot}}, \bibinfo
		{author} {\bibfnamefont {F.}~\bibnamefont {Krausz}}, \bibinfo {author}
		{\bibfnamefont {W.}~\bibnamefont {Yu}}, \bibinfo {author} {\bibfnamefont
			{G.~D.}\ \bibnamefont {Tsakiris}}, \ and\ \bibinfo {author} {\bibfnamefont
			{L.}~\bibnamefont {Veisz}},\ }\href {\doibase 10.1063/1.4914087} {\bibfield
		{journal} {\bibinfo  {journal} {Physics of Plasmas}\ }\textbf {\bibinfo
			{volume} {22}},\ \bibinfo {pages} {033105} (\bibinfo {year}
		{2015})}\BibitemShut {NoStop}%
	\bibitem [{\citenamefont {Kahaly}\ \emph {et~al.}(2013)\citenamefont {Kahaly},
		\citenamefont {Monchoc\'e}, \citenamefont {Vincenti}, \citenamefont
		{Dzelzainis}, \citenamefont {Dromey}, \citenamefont {Zepf}, \citenamefont
		{Martin},\ and\ \citenamefont {Qu\'er\'e}}]{Kahaly13}%
	\BibitemOpen
	\bibfield  {author} {\bibinfo {author} {\bibfnamefont {S.}~\bibnamefont
			{Kahaly}}, \bibinfo {author} {\bibfnamefont {S.}~\bibnamefont {Monchoc\'e}},
		\bibinfo {author} {\bibfnamefont {H.}~\bibnamefont {Vincenti}}, \bibinfo
		{author} {\bibfnamefont {T.}~\bibnamefont {Dzelzainis}}, \bibinfo {author}
		{\bibfnamefont {B.}~\bibnamefont {Dromey}}, \bibinfo {author} {\bibfnamefont
			{M.}~\bibnamefont {Zepf}}, \bibinfo {author} {\bibfnamefont {P.}~\bibnamefont
			{Martin}}, \ and\ \bibinfo {author} {\bibfnamefont {F.}~\bibnamefont
			{Qu\'er\'e}},\ }\href@noop {} {\bibfield  {journal} {\bibinfo  {journal}
			{Phys. Rev. Lett.}\ }\textbf {\bibinfo {volume} {110}},\ \bibinfo {pages}
		{175001} (\bibinfo {year} {2013})}\BibitemShut {NoStop}%
	\bibitem [{\citenamefont {R\"odel}\ \emph {et~al.}(2012)\citenamefont
		{R\"odel}, \citenamefont {an~der Br\"ugge}, \citenamefont {Bierbach},
		\citenamefont {Yeung}, \citenamefont {Hahn}, \citenamefont {Dromey},
		\citenamefont {Herzer}, \citenamefont {Fuchs}, \citenamefont {Pour},
		\citenamefont {Eckner}, \citenamefont {Behmke}, \citenamefont {Cerchez},
		\citenamefont {J\"ackel}, \citenamefont {Hemmers}, \citenamefont {Toncian},
		\citenamefont {Kaluza}, \citenamefont {Belyanin}, \citenamefont {Pretzler},
		\citenamefont {Willi}, \citenamefont {Pukhov}, \citenamefont {Zepf},\ and\
		\citenamefont {Paulus}}]{Jena2012}%
	\BibitemOpen
	\bibfield  {author} {\bibinfo {author} {\bibfnamefont {C.}~\bibnamefont
			{R\"odel}}, \bibinfo {author} {\bibfnamefont {D.}~\bibnamefont {an~der
				Br\"ugge}}, \bibinfo {author} {\bibfnamefont {J.}~\bibnamefont {Bierbach}},
		\bibinfo {author} {\bibfnamefont {M.}~\bibnamefont {Yeung}}, \bibinfo
		{author} {\bibfnamefont {T.}~\bibnamefont {Hahn}}, \bibinfo {author}
		{\bibfnamefont {B.}~\bibnamefont {Dromey}}, \bibinfo {author} {\bibfnamefont
			{S.}~\bibnamefont {Herzer}}, \bibinfo {author} {\bibfnamefont
			{S.}~\bibnamefont {Fuchs}}, \bibinfo {author} {\bibfnamefont {A.~G.}\
			\bibnamefont {Pour}}, \bibinfo {author} {\bibfnamefont {E.}~\bibnamefont
			{Eckner}}, \bibinfo {author} {\bibfnamefont {M.}~\bibnamefont {Behmke}},
		\bibinfo {author} {\bibfnamefont {M.}~\bibnamefont {Cerchez}}, \bibinfo
		{author} {\bibfnamefont {O.}~\bibnamefont {J\"ackel}}, \bibinfo {author}
		{\bibfnamefont {D.}~\bibnamefont {Hemmers}}, \bibinfo {author} {\bibfnamefont
			{T.}~\bibnamefont {Toncian}}, \bibinfo {author} {\bibfnamefont {M.~C.}\
			\bibnamefont {Kaluza}}, \bibinfo {author} {\bibfnamefont {A.}~\bibnamefont
			{Belyanin}}, \bibinfo {author} {\bibfnamefont {G.}~\bibnamefont {Pretzler}},
		\bibinfo {author} {\bibfnamefont {O.}~\bibnamefont {Willi}}, \bibinfo
		{author} {\bibfnamefont {A.}~\bibnamefont {Pukhov}}, \bibinfo {author}
		{\bibfnamefont {M.}~\bibnamefont {Zepf}}, \ and\ \bibinfo {author}
		{\bibfnamefont {G.~G.}\ \bibnamefont {Paulus}},\ }\href {\doibase
		10.1103/PhysRevLett.109.125002} {\bibfield  {journal} {\bibinfo  {journal}
			{Phys. Rev. Lett.}\ }\textbf {\bibinfo {volume} {109}},\ \bibinfo {pages}
		{125002} (\bibinfo {year} {2012})}\BibitemShut {NoStop}%
	\bibitem [{\citenamefont {Dollar}\ \emph {et~al.}(2013)\citenamefont {Dollar},
		\citenamefont {Cummings}, \citenamefont {Chvykov}, \citenamefont
		{Willingale}, \citenamefont {Vargas}, \citenamefont {Yanovsky}, \citenamefont
		{Zulick}, \citenamefont {Maksimchuk}, \citenamefont {Thomas},\ and\
		\citenamefont {Krushelnick}}]{Dollar2013}%
	\BibitemOpen
	\bibfield  {author} {\bibinfo {author} {\bibfnamefont {F.}~\bibnamefont
			{Dollar}}, \bibinfo {author} {\bibfnamefont {P.}~\bibnamefont {Cummings}},
		\bibinfo {author} {\bibfnamefont {V.}~\bibnamefont {Chvykov}}, \bibinfo
		{author} {\bibfnamefont {L.}~\bibnamefont {Willingale}}, \bibinfo {author}
		{\bibfnamefont {M.}~\bibnamefont {Vargas}}, \bibinfo {author} {\bibfnamefont
			{V.}~\bibnamefont {Yanovsky}}, \bibinfo {author} {\bibfnamefont
			{C.}~\bibnamefont {Zulick}}, \bibinfo {author} {\bibfnamefont
			{A.}~\bibnamefont {Maksimchuk}}, \bibinfo {author} {\bibfnamefont {A.~G.~R.}\
			\bibnamefont {Thomas}}, \ and\ \bibinfo {author} {\bibfnamefont
			{K.}~\bibnamefont {Krushelnick}},\ }\href {\doibase
		10.1103/PhysRevLett.110.175002} {\bibfield  {journal} {\bibinfo  {journal}
			{Phys. Rev. Lett.}\ }\textbf {\bibinfo {volume} {110}},\ \bibinfo {pages}
		{175002} (\bibinfo {year} {2013})}\BibitemShut {NoStop}%
	\bibitem [{\citenamefont {Adumi}\ \emph {et~al.}(2004)\citenamefont {Adumi},
		\citenamefont {Tanaka}, \citenamefont {Matsuoka}, \citenamefont {Kurahashi},
		\citenamefont {Yabuuchi}, \citenamefont {Kitagawa}, \citenamefont {Kodama},
		\citenamefont {Sawai}, \citenamefont {Suzuki}, \citenamefont {Okabe},
		\citenamefont {Sera}, \citenamefont {Norimatsu},\ and\ \citenamefont
		{Izawa}}]{plasma_expension2}%
	\BibitemOpen
	\bibfield  {author} {\bibinfo {author} {\bibfnamefont {K.}~\bibnamefont
			{Adumi}}, \bibinfo {author} {\bibfnamefont {K.~A.}\ \bibnamefont {Tanaka}},
		\bibinfo {author} {\bibfnamefont {T.}~\bibnamefont {Matsuoka}}, \bibinfo
		{author} {\bibfnamefont {T.}~\bibnamefont {Kurahashi}}, \bibinfo {author}
		{\bibfnamefont {T.}~\bibnamefont {Yabuuchi}}, \bibinfo {author}
		{\bibfnamefont {Y.}~\bibnamefont {Kitagawa}}, \bibinfo {author}
		{\bibfnamefont {R.}~\bibnamefont {Kodama}}, \bibinfo {author} {\bibfnamefont
			{K.}~\bibnamefont {Sawai}}, \bibinfo {author} {\bibfnamefont
			{K.}~\bibnamefont {Suzuki}}, \bibinfo {author} {\bibfnamefont
			{K.}~\bibnamefont {Okabe}}, \bibinfo {author} {\bibfnamefont
			{T.}~\bibnamefont {Sera}}, \bibinfo {author} {\bibfnamefont {T.}~\bibnamefont
			{Norimatsu}}, \ and\ \bibinfo {author} {\bibfnamefont {Y.}~\bibnamefont
			{Izawa}},\ }\href {\doibase 10.1063/1.1760774} {\bibfield  {journal}
		{\bibinfo  {journal} {Physics of Plasmas}\ }\textbf {\bibinfo {volume}
			{11}},\ \bibinfo {pages} {3721} (\bibinfo {year} {2004})}\BibitemShut
	{NoStop}%
	\bibitem [{\citenamefont {Baltuska}\ \emph {et~al.}(2003)\citenamefont
		{Baltuska}, \citenamefont {Uiberacker}, \citenamefont {Goulielmakis},
		\citenamefont {Kienberger}, \citenamefont {Yakovlev}, \citenamefont {Udem},
		\citenamefont {Hansch},\ and\ \citenamefont {Krausz}}]{CEP_tag}%
	\BibitemOpen
	\bibfield  {author} {\bibinfo {author} {\bibfnamefont {A.}~\bibnamefont
			{Baltuska}}, \bibinfo {author} {\bibfnamefont {M.}~\bibnamefont
			{Uiberacker}}, \bibinfo {author} {\bibfnamefont {E.}~\bibnamefont
			{Goulielmakis}}, \bibinfo {author} {\bibfnamefont {R.}~\bibnamefont
			{Kienberger}}, \bibinfo {author} {\bibfnamefont {V.~S.}\ \bibnamefont
			{Yakovlev}}, \bibinfo {author} {\bibfnamefont {T.}~\bibnamefont {Udem}},
		\bibinfo {author} {\bibfnamefont {T.~W.}\ \bibnamefont {Hansch}}, \ and\
		\bibinfo {author} {\bibfnamefont {F.}~\bibnamefont {Krausz}},\ }\href@noop {}
	{\bibfield  {journal} {\bibinfo  {journal} {IEEE Journal of Selected Topics
				in Quantum Electronics}\ }\textbf {\bibinfo {volume} {9}},\ \bibinfo {pages}
		{972} (\bibinfo {year} {2003})}\BibitemShut {NoStop}%
	\bibitem [{\citenamefont {Li}\ \emph {et~al.}(2007)\citenamefont {Li},
		\citenamefont {Moon}, \citenamefont {Wang}, \citenamefont {Mashiko},
		\citenamefont {Nakamura}, \citenamefont {Tackett},\ and\ \citenamefont
		{Chang}}]{f2f}%
	\BibitemOpen
	\bibfield  {author} {\bibinfo {author} {\bibfnamefont {C.}~\bibnamefont
			{Li}}, \bibinfo {author} {\bibfnamefont {E.}~\bibnamefont {Moon}}, \bibinfo
		{author} {\bibfnamefont {H.}~\bibnamefont {Wang}}, \bibinfo {author}
		{\bibfnamefont {H.}~\bibnamefont {Mashiko}}, \bibinfo {author} {\bibfnamefont
			{C.~M.}\ \bibnamefont {Nakamura}}, \bibinfo {author} {\bibfnamefont
			{J.}~\bibnamefont {Tackett}}, \ and\ \bibinfo {author} {\bibfnamefont
			{Z.}~\bibnamefont {Chang}},\ }\href {\doibase 10.1364/OL.32.000796}
	{\bibfield  {journal} {\bibinfo  {journal} {Opt. Lett.}\ }\textbf {\bibinfo
			{volume} {32}},\ \bibinfo {pages} {796} (\bibinfo {year} {2007})}\BibitemShut
	{NoStop}%
	\bibitem [{sup()}]{sup_mat}%
	\BibitemOpen
	\href@noop {} {\bibinfo  {journal} {See Supplemental Material at [URL will be
			inserted by publisher] for more details on the derivation of the equations
			and data processing}\ }\BibitemShut {NoStop}%
	\bibitem [{\citenamefont {Lichters}\ \emph {et~al.}(1996)\citenamefont
		{Lichters}, \citenamefont {Meyer-ter Vehn},\ and\ \citenamefont
		{Pukhov}}]{LPIC}%
	\BibitemOpen
	\bibfield  {journal} {  }\bibfield  {author} {\bibinfo {author} {\bibfnamefont
			{R.}~\bibnamefont {Lichters}}, \bibinfo {author} {\bibfnamefont
			{J.}~\bibnamefont {Meyer-ter Vehn}}, \ and\ \bibinfo {author} {\bibfnamefont
			{A.}~\bibnamefont {Pukhov}},\ }\href {\doibase 10.1063/1.871619} {\bibfield
		{journal} {\bibinfo  {journal} {Physics of Plasmas}\ }\textbf {\bibinfo
			{volume} {3}},\ \bibinfo {pages} {3425} (\bibinfo {year} {1996})}\BibitemShut
	{NoStop}%
	\bibitem [{\citenamefont {Borot}\ \emph {et~al.}(2012)\citenamefont {Borot},
		\citenamefont {Malvache}, \citenamefont {Chen}, \citenamefont {Jullien},
		\citenamefont {Geindre}, \citenamefont {Audebert}, \citenamefont {Mourou},
		\citenamefont {Qu{\'e}r{\'e}},\ and\ \citenamefont
		{Lopez-Martens}}]{Borot2012}%
	\BibitemOpen
	\bibfield  {author} {\bibinfo {author} {\bibfnamefont {A.}~\bibnamefont
			{Borot}}, \bibinfo {author} {\bibfnamefont {A.}~\bibnamefont {Malvache}},
		\bibinfo {author} {\bibfnamefont {X.}~\bibnamefont {Chen}}, \bibinfo {author}
		{\bibfnamefont {A.}~\bibnamefont {Jullien}}, \bibinfo {author} {\bibfnamefont
			{J.-P.}\ \bibnamefont {Geindre}}, \bibinfo {author} {\bibfnamefont
			{P.}~\bibnamefont {Audebert}}, \bibinfo {author} {\bibfnamefont
			{G.}~\bibnamefont {Mourou}}, \bibinfo {author} {\bibfnamefont
			{F.}~\bibnamefont {Qu{\'e}r{\'e}}}, \ and\ \bibinfo {author} {\bibfnamefont
			{R.}~\bibnamefont {Lopez-Martens}},\ }\href
	{http://dx.doi.org/10.1038/nphys2269} {\bibfield  {journal} {\bibinfo
			{journal} {Nature Physics}\ }\textbf {\bibinfo {volume} {8}},\ \bibinfo
		{pages} {416} (\bibinfo {year} {2012})}\BibitemShut {NoStop}%
	\bibitem [{\citenamefont {an~der Brügge}\ and\ \citenamefont
		{Pukhov}(2010)}]{CSE_Pukhov}%
	\BibitemOpen
	\bibfield  {author} {\bibinfo {author} {\bibfnamefont {D.}~\bibnamefont
			{an~der Brügge}}\ and\ \bibinfo {author} {\bibfnamefont {A.}~\bibnamefont
			{Pukhov}},\ }\href {\doibase 10.1063/1.3353050} {\bibfield  {journal}
		{\bibinfo  {journal} {Physics of Plasmas}\ }\textbf {\bibinfo {volume}
			{17}},\ \bibinfo {pages} {033110} (\bibinfo {year} {2010})}\BibitemShut
	{NoStop}%
	\bibitem [{\citenamefont {Qu\'er\'e}\ \emph {et~al.}(2006)\citenamefont
		{Qu\'er\'e}, \citenamefont {Thaury}, \citenamefont {Monot}, \citenamefont
		{Dobosz}, \citenamefont {Martin}, \citenamefont {Geindre},\ and\
		\citenamefont {Audebert}}]{CWE}%
	\BibitemOpen
	\bibfield  {author} {\bibinfo {author} {\bibfnamefont {F.}~\bibnamefont
			{Qu\'er\'e}}, \bibinfo {author} {\bibfnamefont {C.}~\bibnamefont {Thaury}},
		\bibinfo {author} {\bibfnamefont {P.}~\bibnamefont {Monot}}, \bibinfo
		{author} {\bibfnamefont {S.}~\bibnamefont {Dobosz}}, \bibinfo {author}
		{\bibfnamefont {P.}~\bibnamefont {Martin}}, \bibinfo {author} {\bibfnamefont
			{J.-P.}\ \bibnamefont {Geindre}}, \ and\ \bibinfo {author} {\bibfnamefont
			{P.}~\bibnamefont {Audebert}},\ }\href {\doibase
		10.1103/PhysRevLett.96.125004} {\bibfield  {journal} {\bibinfo  {journal}
			{Phys. Rev. Lett.}\ }\textbf {\bibinfo {volume} {96}},\ \bibinfo {pages}
		{125004} (\bibinfo {year} {2006})}\BibitemShut {NoStop}%
	\bibitem [{\citenamefont {Vincenti}\ \emph {et~al.}(2013)\citenamefont
		{Vincenti}, \citenamefont {Monchocé}, \citenamefont {Kahaly}, \citenamefont
		{Bonnaud}, \citenamefont {Martin},\ and\ \citenamefont {Quéré}}]{denting}%
	\BibitemOpen
	\bibfield  {author} {\bibinfo {author} {\bibfnamefont {H.}~\bibnamefont
			{Vincenti}}, \bibinfo {author} {\bibfnamefont {S.}~\bibnamefont {Monchocé}},
		\bibinfo {author} {\bibfnamefont {S.}~\bibnamefont {Kahaly}}, \bibinfo
		{author} {\bibfnamefont {G.}~\bibnamefont {Bonnaud}}, \bibinfo {author}
		{\bibfnamefont {P.}~\bibnamefont {Martin}}, \ and\ \bibinfo {author}
		{\bibfnamefont {F.}~\bibnamefont {Quéré}},\ }\href@noop {} {\bibfield
		{journal} {\bibinfo  {journal} {Nature Communications}\ }\textbf {\bibinfo
			{volume} {5}},\ \bibinfo {pages} {3403} (\bibinfo {year} {2013})}\BibitemShut
	{NoStop}%
	\bibitem [{\citenamefont {Gonoskov}(2018)}]{Gonoskov2018}%
	\BibitemOpen
	\bibfield  {author} {\bibinfo {author} {\bibfnamefont {A.}~\bibnamefont
			{Gonoskov}},\ }\href {\doibase 10.1063/1.5000785} {\bibfield  {journal}
		{\bibinfo  {journal} {Physics of Plasmas}\ }\textbf {\bibinfo {volume}
			{25}},\ \bibinfo {pages} {013108} (\bibinfo {year} {2018})}\BibitemShut
	{NoStop}%
	\bibitem [{\citenamefont {Gonoskov}\ \emph {et~al.}(2011)\citenamefont
		{Gonoskov}, \citenamefont {Korzhimanov}, \citenamefont {Kim}, \citenamefont
		{Marklund},\ and\ \citenamefont {Sergeev}}]{Gonoskov2011}%
	\BibitemOpen
	\bibfield  {author} {\bibinfo {author} {\bibfnamefont {A.~A.}\ \bibnamefont
			{Gonoskov}}, \bibinfo {author} {\bibfnamefont {A.~V.}\ \bibnamefont
			{Korzhimanov}}, \bibinfo {author} {\bibfnamefont {A.~V.}\ \bibnamefont
			{Kim}}, \bibinfo {author} {\bibfnamefont {M.}~\bibnamefont {Marklund}}, \
		and\ \bibinfo {author} {\bibfnamefont {A.~M.}\ \bibnamefont {Sergeev}},\
	}\href {\doibase 10.1103/PhysRevE.84.046403} {\bibfield  {journal} {\bibinfo
			{journal} {Phys. Rev. E}\ }\textbf {\bibinfo {volume} {84}},\ \bibinfo
		{pages} {046403} (\bibinfo {year} {2011})}\BibitemShut {NoStop}%
	\bibitem [{\citenamefont {Drescher}\ \emph {et~al.}(2001)\citenamefont
		{Drescher}, \citenamefont {Hentschel}, \citenamefont {Kienberger},
		\citenamefont {Tempea}, \citenamefont {Spielmann}, \citenamefont {Reider},
		\citenamefont {Corkum},\ and\ \citenamefont {Krausz}}]{Drescher2001}%
	\BibitemOpen
	\bibfield  {author} {\bibinfo {author} {\bibfnamefont {M.}~\bibnamefont
			{Drescher}}, \bibinfo {author} {\bibfnamefont {M.}~\bibnamefont {Hentschel}},
		\bibinfo {author} {\bibfnamefont {R.}~\bibnamefont {Kienberger}}, \bibinfo
		{author} {\bibfnamefont {G.}~\bibnamefont {Tempea}}, \bibinfo {author}
		{\bibfnamefont {C.}~\bibnamefont {Spielmann}}, \bibinfo {author}
		{\bibfnamefont {G.~A.}\ \bibnamefont {Reider}}, \bibinfo {author}
		{\bibfnamefont {P.~B.}\ \bibnamefont {Corkum}}, \ and\ \bibinfo {author}
		{\bibfnamefont {F.}~\bibnamefont {Krausz}},\ }\href {\doibase
		10.1126/science.1058561} {\ \textbf {\bibinfo {volume} {291}},\ \bibinfo
		{pages} {1923} (\bibinfo {year} {2001})}\BibitemShut {NoStop}%
	\bibitem [{\citenamefont {Itatani}\ \emph {et~al.}(2002)\citenamefont
		{Itatani}, \citenamefont {Qu\'er\'e}, \citenamefont {Yudin}, \citenamefont
		{Ivanov}, \citenamefont {Krausz},\ and\ \citenamefont
		{Corkum}}]{Itatani2002}%
	\BibitemOpen
	\bibfield  {author} {\bibinfo {author} {\bibfnamefont {J.}~\bibnamefont
			{Itatani}}, \bibinfo {author} {\bibfnamefont {F.}~\bibnamefont {Qu\'er\'e}},
		\bibinfo {author} {\bibfnamefont {G.~L.}\ \bibnamefont {Yudin}}, \bibinfo
		{author} {\bibfnamefont {M.~Y.}\ \bibnamefont {Ivanov}}, \bibinfo {author}
		{\bibfnamefont {F.}~\bibnamefont {Krausz}}, \ and\ \bibinfo {author}
		{\bibfnamefont {P.~B.}\ \bibnamefont {Corkum}},\ }\href {\doibase
		10.1103/PhysRevLett.88.173903} {\bibfield  {journal} {\bibinfo  {journal}
			{Phys. Rev. Lett.}\ }\textbf {\bibinfo {volume} {88}},\ \bibinfo {pages}
		{173903} (\bibinfo {year} {2002})}\BibitemShut {NoStop}%
	\bibitem [{\citenamefont {Kienberger}\ \emph {et~al.}(2004)\citenamefont
		{Kienberger}, \citenamefont {Goulielmakis}, \citenamefont {Uiberacker},
		\citenamefont {Baltuska}, \citenamefont {Yakovlev}, \citenamefont {Bammer},
		\citenamefont {Scrinzi}, \citenamefont {Westerwalbesloh}, \citenamefont
		{Kleineberg}, \citenamefont {Heinzmann}, \citenamefont {Drescher},\ and\
		\citenamefont {Krausz}}]{Kienberger2004}%
	\BibitemOpen
	\bibfield  {author} {\bibinfo {author} {\bibfnamefont {R.}~\bibnamefont
			{Kienberger}}, \bibinfo {author} {\bibfnamefont {E.}~\bibnamefont
			{Goulielmakis}}, \bibinfo {author} {\bibfnamefont {M.}~\bibnamefont
			{Uiberacker}}, \bibinfo {author} {\bibfnamefont {A.}~\bibnamefont
			{Baltuska}}, \bibinfo {author} {\bibfnamefont {V.}~\bibnamefont {Yakovlev}},
		\bibinfo {author} {\bibfnamefont {F.}~\bibnamefont {Bammer}}, \bibinfo
		{author} {\bibfnamefont {A.}~\bibnamefont {Scrinzi}}, \bibinfo {author}
		{\bibfnamefont {T.}~\bibnamefont {Westerwalbesloh}}, \bibinfo {author}
		{\bibfnamefont {U.}~\bibnamefont {Kleineberg}}, \bibinfo {author}
		{\bibfnamefont {U.}~\bibnamefont {Heinzmann}}, \bibinfo {author}
		{\bibfnamefont {M.}~\bibnamefont {Drescher}}, \ and\ \bibinfo {author}
		{\bibfnamefont {F.}~\bibnamefont {Krausz}},\ }\href
	{http://dx.doi.org/10.1038/nature02277} {\bibfield  {journal} {\bibinfo
			{journal} {Nature}\ }\textbf {\bibinfo {volume} {427}},\ \bibinfo {pages}
		{817} (\bibinfo {year} {2004})}\BibitemShut {NoStop}%
	\bibitem [{\citenamefont {Yakovlev}\ \emph {et~al.}(2005)\citenamefont
		{Yakovlev}, \citenamefont {Bammer},\ and\ \citenamefont
		{Scrinzi}}]{Yakovlev05}%
	\BibitemOpen
	\bibfield  {author} {\bibinfo {author} {\bibfnamefont {V.~S.}\ \bibnamefont
			{Yakovlev}}, \bibinfo {author} {\bibfnamefont {F.}~\bibnamefont {Bammer}}, \
		and\ \bibinfo {author} {\bibfnamefont {A.}~\bibnamefont {Scrinzi}},\ }\href
	{\doibase 10.1080/09500340412331283642} {\bibfield  {journal} {\bibinfo
			{journal} {Journal of Modern Optics}\ }\textbf {\bibinfo {volume} {52}},\
		\bibinfo {pages} {395} (\bibinfo {year} {2005})}\BibitemShut {NoStop}%
	\bibitem [{\citenamefont {Budri\={u}nas}\ \emph {et~al.}(2017)\citenamefont
		{Budri\={u}nas}, \citenamefont {Stanislauskas}, \citenamefont {Adamonis},
		\citenamefont {Aleknavi\v{c}ius}, \citenamefont {Veitas}, \citenamefont
		{Gadonas}, \citenamefont {Balickas}, \citenamefont {Michailovas},\ and\
		\citenamefont {Varanavi\v{c}ius}}]{Budriunas:17}%
	\BibitemOpen
	\bibfield  {author} {\bibinfo {author} {\bibfnamefont {R.}~\bibnamefont
			{Budri\={u}nas}}, \bibinfo {author} {\bibfnamefont {T.}~\bibnamefont
			{Stanislauskas}}, \bibinfo {author} {\bibfnamefont {J.}~\bibnamefont
			{Adamonis}}, \bibinfo {author} {\bibfnamefont {A.}~\bibnamefont
			{Aleknavi\v{c}ius}}, \bibinfo {author} {\bibfnamefont {G.}~\bibnamefont
			{Veitas}}, \bibinfo {author} {\bibfnamefont {D.}~\bibnamefont {Gadonas}},
		\bibinfo {author} {\bibfnamefont {S.}~\bibnamefont {Balickas}}, \bibinfo
		{author} {\bibfnamefont {A.}~\bibnamefont {Michailovas}}, \ and\ \bibinfo
		{author} {\bibfnamefont {A.}~\bibnamefont {Varanavi\v{c}ius}},\ }\href
	{\doibase 10.1364/OE.25.005797} {\bibfield  {journal} {\bibinfo  {journal}
			{Opt. Express}\ }\textbf {\bibinfo {volume} {25}},\ \bibinfo {pages} {5797}
		(\bibinfo {year} {2017})}\BibitemShut {NoStop}%
	\bibitem [{\citenamefont {Tzallas}\ \emph {et~al.}(2003)\citenamefont
		{Tzallas}, \citenamefont {Charalambidis}, \citenamefont {Papadogiannis},
		\citenamefont {Witte},\ and\ \citenamefont {Tsakiris}}]{Tzallas2003}%
	\BibitemOpen
	\bibfield  {author} {\bibinfo {author} {\bibfnamefont {P.}~\bibnamefont
			{Tzallas}}, \bibinfo {author} {\bibfnamefont {D.}~\bibnamefont
			{Charalambidis}}, \bibinfo {author} {\bibfnamefont {N.~A.}\ \bibnamefont
			{Papadogiannis}}, \bibinfo {author} {\bibfnamefont {K.}~\bibnamefont
			{Witte}}, \ and\ \bibinfo {author} {\bibfnamefont {G.~D.}\ \bibnamefont
			{Tsakiris}},\ }\href {http://dx.doi.org/10.1038/nature02091} {\bibfield
		{journal} {\bibinfo  {journal} {Nature}\ }\textbf {\bibinfo {volume} {426}},\
		\bibinfo {pages} {267} (\bibinfo {year} {2003})}\BibitemShut {NoStop}%
\end{thebibliography}
\end{document}

	%	\preprint{APS/123-QED}
	
	\title[]{Supplementary Material for \textquotedblleft Towards temporal characterization of intense isolated attosecond pulses from relativistic surface high harmonics\textquotedblright.}
	
	\author{O. Jahn}
	\affiliation{Max-Planck-Institut f\"ur Quantenoptik, 85748 Garching, Germany}
	\affiliation{Department f\"ur Physik, Ludwig-Maximilians-Universit\"at M\"unchen, 85748 Garching, Germany}
	
	\author{V. E. Leshchenko}
	%	\email{vyacheslav.leshchenko@mpq.mpg.de}
	\affiliation{Max-Planck-Institut f\"ur Quantenoptik, 85748 Garching, Germany}
	\affiliation{Department f\"ur Physik, Ludwig-Maximilians-Universit\"at M\"unchen, 85748 Garching, Germany}
	
	\author{P. Tzallas}
	\affiliation{Foundation for Research and Technology-Hellas, Institute of Electronic Structure and Laser, Heraklion, Crete, Greece}
	\affiliation{ELI-ALPS, ELI-HU Non-Profit Ltd., Dugonics t\'er 13, Szeged 6720, Hungary} 
	
	\author{A. Kessel}
	\affiliation{Max-Planck-Institut f\"ur Quantenoptik, 85748 Garching, Germany}
	\affiliation{Department f\"ur Physik, Ludwig-Maximilians-Universit\"at M\"unchen, 85748 Garching, Germany}
	
	\author{M. Kr\"uger}
	\affiliation{Max-Planck-Institut f\"ur Quantenoptik, 85748 Garching, Germany}
	\affiliation{Department f\"ur Physik, Ludwig-Maximilians-Universit\"at M\"unchen, 85748 Garching, Germany}
	
	\author{A. M\"unzer}
	\affiliation{Max-Planck-Institut f\"ur Quantenoptik, 85748 Garching, Germany}
	\affiliation{Department f\"ur Physik, Ludwig-Maximilians-Universit\"at M\"unchen, 85748 Garching, Germany}
	
	\author{S. A. Trushin}
	\affiliation{Max-Planck-Institut f\"ur Quantenoptik, 85748 Garching, Germany}
	\affiliation{Department f\"ur Physik, Ludwig-Maximilians-Universit\"at M\"unchen, 85748 Garching, Germany}
	
	\author{M. Schultze}
	\affiliation{Max-Planck-Institut f\"ur Quantenoptik, 85748 Garching, Germany}
	
	\author{G. D. Tsakiris}
	\affiliation{Max-Planck-Institut f\"ur Quantenoptik, 85748 Garching, Germany}
	
	\author{S. Kahaly}
	\affiliation{ELI-ALPS, ELI-HU Non-Profit Ltd., Dugonics t\'er 13, Szeged 6720, Hungary}    
	
	\author{A. Guggenmos}
	\affiliation{Max-Planck-Institut f\"ur Quantenoptik, 85748 Garching, Germany}
	\affiliation{Department f\"ur Physik, Ludwig-Maximilians-Universit\"at M\"unchen, 85748 Garching, Germany}
	
	\author{D. Kormin}
	\affiliation{Max-Planck-Institut f\"ur Quantenoptik, 85748 Garching, Germany}
	\affiliation{Department f\"ur Physik, Ludwig-Maximilians-Universit\"at M\"unchen, 85748 Garching, Germany}
	
	\author{L. Veisz}
	\affiliation{Max-Planck-Institut f\"ur Quantenoptik, 85748 Garching, Germany}
	\affiliation{Department of Physics, Ume\r{a} University, Ume\r{a}, Sweden}
	
	\author{F. Krausz}
	\affiliation{Max-Planck-Institut f\"ur Quantenoptik, 85748 Garching, Germany}
	\affiliation{Department f\"ur Physik, Ludwig-Maximilians-Universit\"at M\"unchen, 85748 Garching, Germany}
	
	\author{Zs. Major}
	\affiliation{Max-Planck-Institut f\"ur Quantenoptik, 85748 Garching, Germany}
	\affiliation{Department f\"ur Physik, Ludwig-Maximilians-Universit\"at M\"unchen, 85748 Garching, Germany}
	
	\author{S. Karsch}
	%	\email{stefan.karsch@mpq.mpg.de}
	\affiliation{Max-Planck-Institut f\"ur Quantenoptik, 85748 Garching, Germany}
	\affiliation{Department f\"ur Physik, Ludwig-Maximilians-Universit\"at M\"unchen, 85748 Garching, Germany}

	%	\date{\today}
	\maketitle
	
	\section{Data acquisition and processing for CEP dependence}
	
	In this section we provide the technical details on the data acquisition and processing leading to the CEP dependence of the measured harmonics.
	Experimental realization of a CEP scan in a low-repetition-rate system with not perfectly stable laser pulse parameters is a challenging but feasible task. The common approach to overcome such a problem is to log all parameters of the laser system and select the shots with an acceptable deviation from the mean values in the later analysis. In our case, the root mean squared (RMS) energy instability of the driving pulses was 4\% which can cause a significant error in the f-2f measurements. To exclude noise and uncertainties originating from system instabilities, OPCPA energy and spectrum were logged in parallel to the harmonic spectrum and f-2f signal.
	%In particular, the leakage through the specially designed silver mirror with about 0.3\% transmission installed on the compressor output was measured.
	Afterwards, the shots for which the OPCPA energy and central wavelength deviated by not more than 1\% from the mean values were selected which, in particular, limits the error of the f-2f interferometer originating from input energy instability to less than $\sim$200\,mrad [20]. After this sorting procedure about 20\% of the collected data were left for the final analysis.
	
	Finally, each shot was assigned with a CEP value according to the measured f-2f phase data. To average over the target instabilities and other uncontrolled generation conditions, the full CEP-range of 0--2$\pi$ was divided into 12 intervals of equal size into which the shots were sorted. The spectra averaged over each interval are shown in Fig.~2(a) in the main paper.

	\section{Temporal delay variation versus harmonic position shift}
	
	In this we will give an estimation of the averaged temporal delay variation of the attosecond pulse train resulting from the shift of the harmonic position in the detected spectrum. A train of pulses separated by a delay of $T$ in the time domain results in spectral modulations in the frequency domain with a separation of $F=1/T$ between the maxima (which we will call in the following \textquotedblleft harmonics\textquotedblright). The frequency of the n-th harmonic is then $F_n = n\times F$.	When the delay T changes by $\Delta T$, the frequency spacing F between the harmonics changes accordingly by $\Delta F=1/T-1/(T+\Delta T)\approx\Delta T/T^2$. For the n-th harmonic this results in a spectral shift of $\Delta F_n=n\times \Delta F\approx n\times \Delta T/T^2$. Conversely, a frequency shift of the harmonics can be related to a change in the time delay. In the particular case presented in the paper when the harmonic position is shifted by one harmonic order, the change of the delay can be derived from the following relation:
	\begin{math}
	n/T=(n+1)/(T+\Delta T),
	\end{math}
	which yields the equation (1) used in the paper
	\begin{equation}
	\Delta T=T/n.
	\end{equation}
	
	\section{Denting estimation from the spectral beating period}	
	
	Let us now develop the above scenario further and consider a pulse train with an uneven spacing between pulses. In the simplest case with three pulses and a delay of $T_1=T$ between 1$^{\rm st}$ and 2$^{\rm nd}$ pulses and a delay of $T_2=T+\delta T$ between 2$^{\rm nd}$ and 3$^{\rm rd}$ pulses, the spectral intensity is given by
	
	\begin{align}
		I(f) &= I_0(f)\times \Big|1+e^{i2\pi fT_1}+e^{i2\pi fT_2}\Big|^2 \nonumber\\
		&= I_0(f)\times \Big(3+2\cos\big(2\pi fT\big) + \nonumber\\
		&\hspace{0.5cm} +2\cos\big(2\pi f(T+\delta T)\big)+2\cos\big(2\pi f\delta T\big)\Big) \nonumber\\ 
		&\approx I_0(f)\times \Big(3+2\cos\big(2\pi fT\big) + \nonumber\\
		&\hspace{0.5cm} + 2\cos\big(2\pi fT\big)\cos\big(2\pi f\delta T\big)+2\cos\big(2\pi f\delta T\big)\Big) \nonumber\\
		&=I_0(f)\times\Big(1+8\cos^2\big(\pi fT\big)\cos^2\big(\pi f\delta T\big)\Big)
	\end{align}
	
	with $f$ being the frequency. An identical spectrum $I_0(f)$ is assumed for all three pulses stacked with the above defined time delays. In this equation the first cosine-factor can be identified as the previously described harmonic modulation of the spectrum, while the second factor further modulates this spectrum with a beating period of $f_{\text{beating}}=1/\delta T$. For a given beating period, e.g.\ as extracted from our experimental data, the delay variation can be determined by
	\begin{equation}
	\delta T=1/f_{\text{beating}}.
	\end{equation}
	
\begin{figure}[t]
		
		\includegraphics[width=0.9\linewidth]{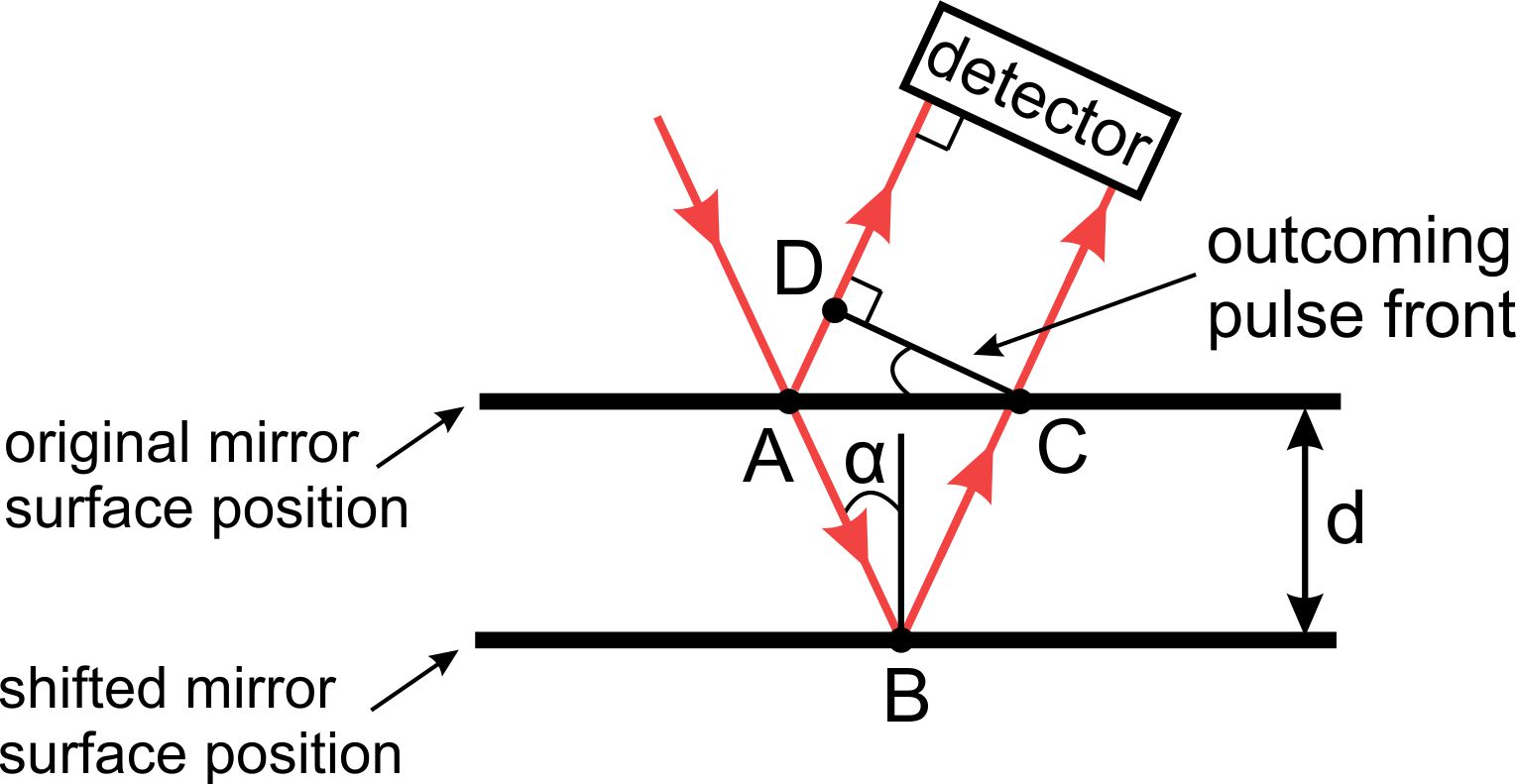}
		\caption{Sketch illustrating the altered optical path lengths for a shifted plasma mirror surface (\textquotedblleft denting\textquotedblright).}
		\label{fig:denting}
	\end{figure}

	As the next step, we can use this delay variation to estimate the plasma denting, i.e.\ the relative shift of the reflection point from the plasma mirror between the different events of attosecond pulse generation along the pulse train. As illustrated in Fig.~\ref{fig:denting}, a shift of the mirror surface by $d$ results in an additional optical path length $\Delta L$ which is geometrically determined by
	
	\begin{math}
	\Delta L=AB+BC-AD=\\=2d/\cos(\alpha)-2d\tan(\alpha)\sin(\alpha)=2d\cos(\alpha),
	\end{math}
	
	where $\alpha$ is the angle of incidence. Solving this equation for $d$ and replacing $\Delta L$ with the delay variation $\delta T$ times the speed of light $c$ results in the equation (\ref{d}) used in the paper for the denting estimation:
	\begin{equation}
	\label{d}
	d=\Delta L/(2\cos(\alpha))=c\times \delta T/(2\cos(\alpha)).
	\end{equation}

	\section{CEP dependence of the harmonic spectrum and comparison with theory}
	
	Here we give details on the origin of the CEP dependence of the harmonic spectrum and put our experimental data into perspective with the established models of relativistic SHHG.
	
	\begin{figure}[]
		\includegraphics[width=\linewidth]{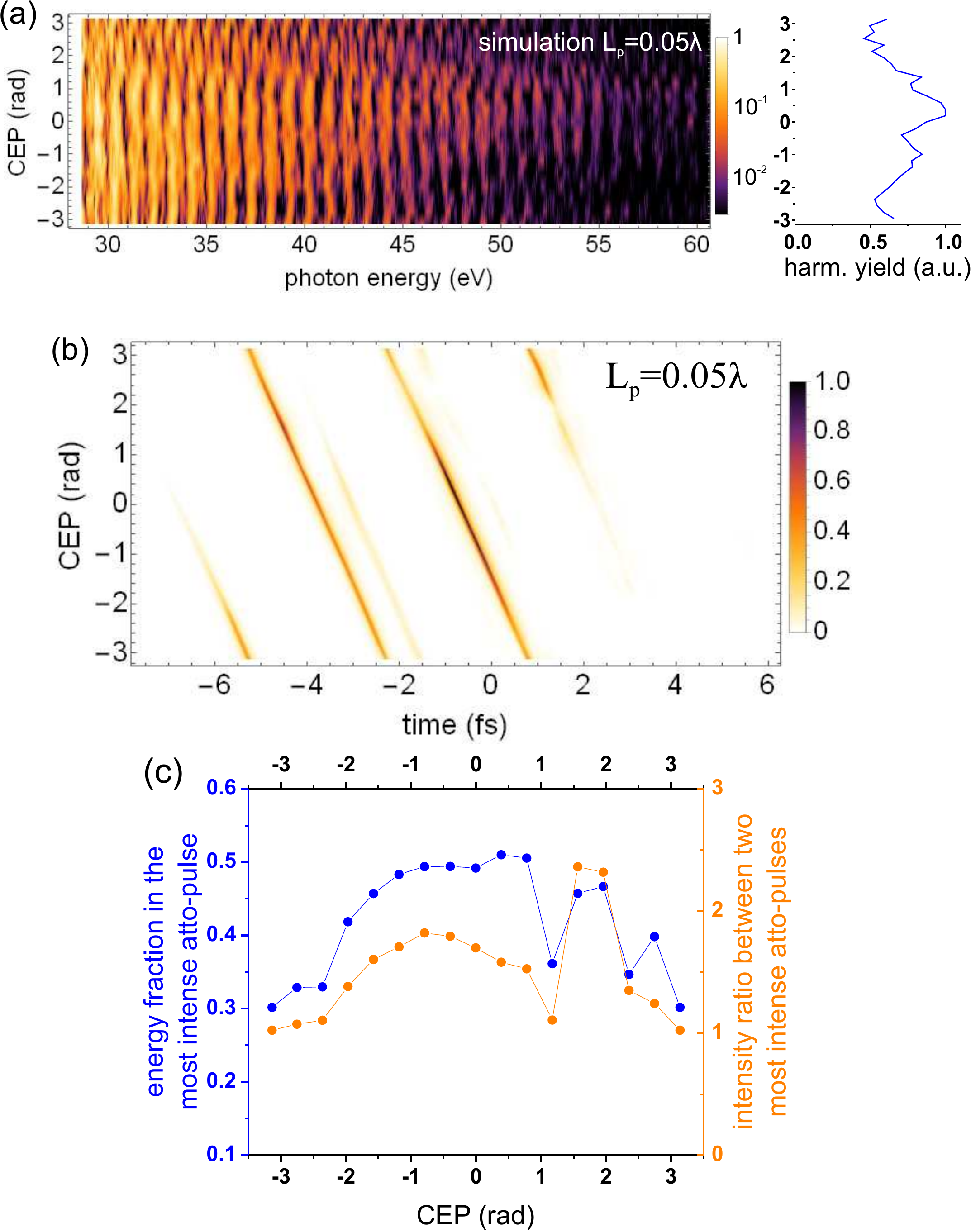}
		\caption{\textbf{(a)} PIC simulations of the harmonic spectrum (left panels) and the integrated harmonic yield (right panels) on the CEP of the driving field for $L_p=0.05\lambda$. \textbf{(b)} CEP dependence of the temporal structure of the attosecond pulse train for $L_p=0.05\lambda$ (200 nm thick Al filter is applied). \textbf{(c)} Energy (blue-dotted line) and intensity (orange-dotted line) ratio between the main attosecond pulse and the rest of the train. In these simulations $a_0=3$, $\tau=7\thinspace \text{fs}$, $L_p=0.05\lambda$.}
		\label{fig:temp005}
	\end{figure}
	
\begin{figure}[]
		\includegraphics[width=\linewidth]{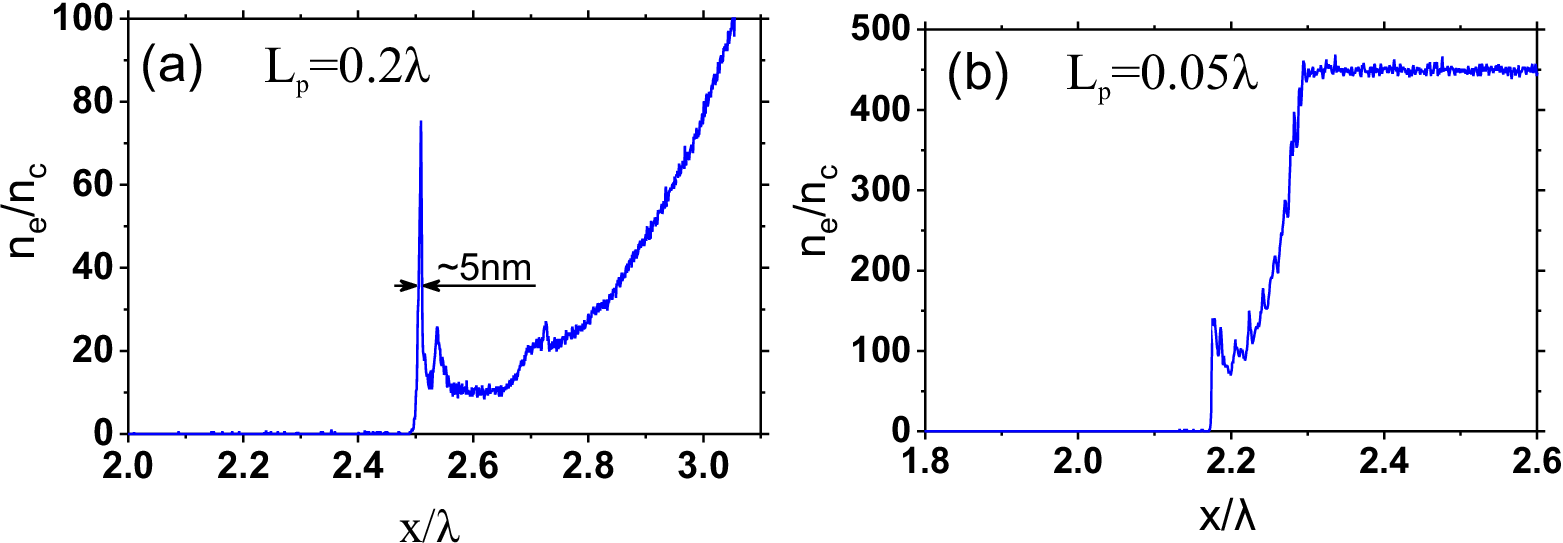}
		\caption{Plasma electron density right before the generation of the most intense attosecond pulse for CEP=-0.32 rad, $L_p=0.2\lambda$ (a) (temporal structure is presented in the paper in Fig.4) and $L_p=0.05\lambda$ (b).}
		\label{fig:ebuch02}
	\end{figure}

	As written in the main text of the paper, there are several publications that describe the basic mechanisms underlying the harmonics CEP-shift and the generation of an isolated attosecond pulse [14, 24, 27, 28] which also predict the key importance of the plasma scale length optimization. In order to investigate the influence of the plasma scale length on the CEP-dependence of spectral and temporal structures of the generated harmonics, we performed a PIC simulation for $L_p=0.05\lambda$, i.e.\ a smaller scale length compared to the case of $L_p=0.2\lambda$ presented in the main text, at otherwise identical laser parameters. The result is displayed in Fig.~\ref{fig:temp005} and shows no CEP dependence of the harmonic spectra for $L_p=0.05\lambda$ and the generation of two nearly equal attosecond pulses for all CEP values. The comparison of these results with the simulations for $L_p=0.2\lambda$ presented in the paper demonstrates the crucial importance of the plasma scale length optimization for the generation of isolated attosecond pulses. In particular the analysis of the plasma dynamics in PIC simulations reveals clear differences in the electron plasma density. This is shown in Fig.~\ref{fig:ebuch02}, which depicts the electron density right before the generation of the most intense attosecond pulse for two different plasma scale lengths. While for $L_p = 0.2\lambda$ an electron bunch with 5\,nm layer thickness is created, no such feature is present for $L_p= 0.05\lambda$. This can be related to the steeper density gradient in the latter case where electrons at the plasma edge are just pushed into the bulk rather than bunched into a thin layer. A more detailed analysis of this effect can be found in [27]. The simulations show that for $L_p = 0.2\lambda$ the position of the electron nano-bunch shifts deeper inside the plasma with every subsequent optical cycle. This effect is known as plasma denting [26] and causes an increase of the temporal spacing between the pulses in the generated attosecond pulse train. Since the magnitude of each shift is determined by the field strength of the previous optical cycle which depends on the CEP, there is a dependence of the attosecond pulse spacing and thus the harmonic spectral structure on the CEP of the driving field. (The connection between the pulse temporal spacing and the harmonic spectral structure was discussed above in section II).
	
\begin{figure}[]
		\includegraphics[width=0.7\linewidth]{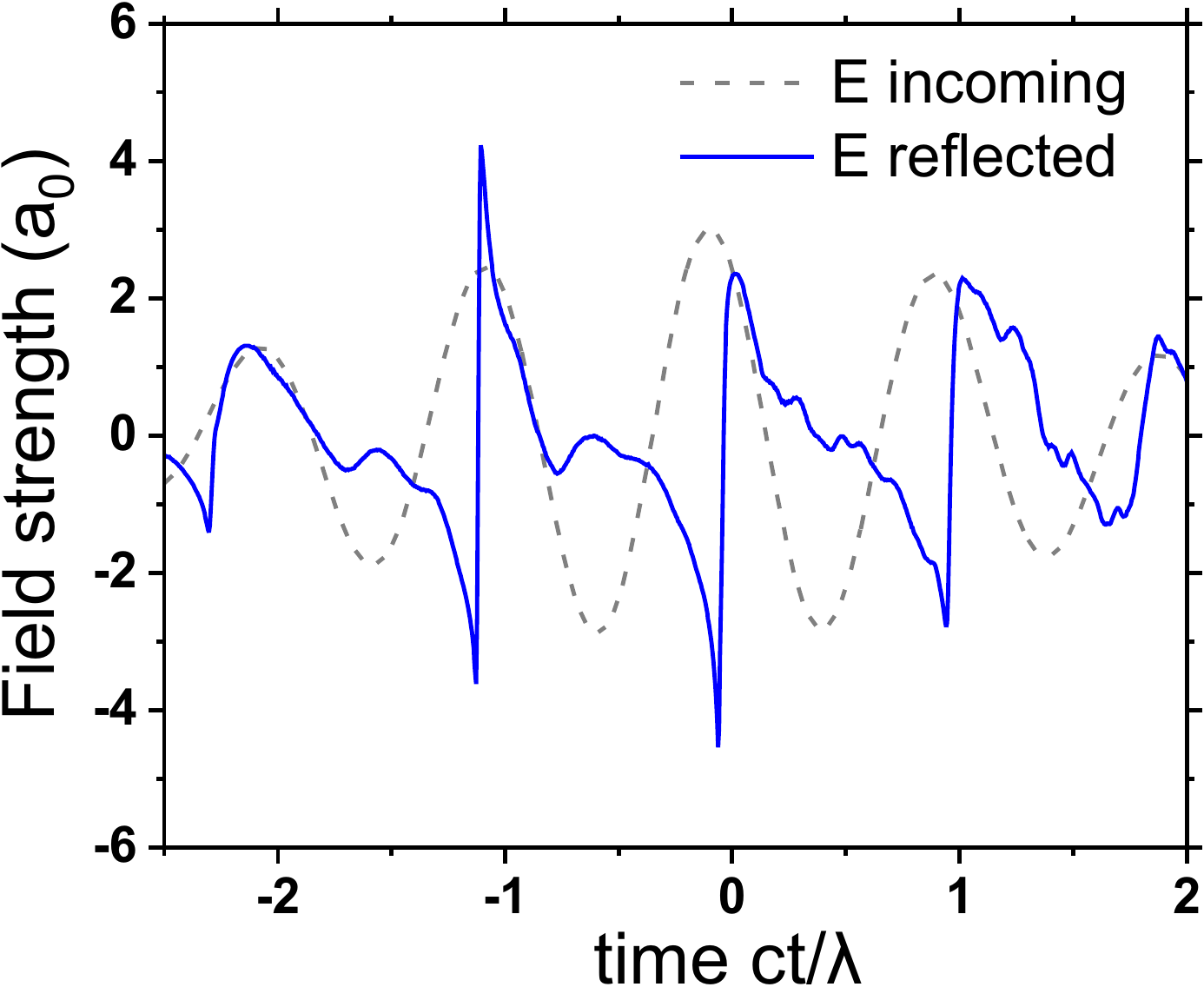}
		\caption{Incoming and reflected from the plasma mirror electric field. In these simulations $a_0=3$, $\tau=7\thinspace \text{fs}$, $L_p=0.2\lambda$, CEP=-0.32\,rad.}
		\label{fig:refE}
	\end{figure}

	A further analysis of the results from our PIC simulation at $L_p = 0.2\lambda$ reveals that the amplitude of the reflected field exceeds the incoming one by up to 50\% (see Fig.~\ref{fig:refE}). This effect requires a temporary storage of energy and cannot be explained within the relativistic oscillating mirror (ROM) model [10]. However, such a mechanism exists in the coherent synchrotron emission (CSE) [24] and the relativistic electronic spring (RES) model [28] in the form of the already mentioned electron nano-bunch: via the charge separation between the electrons and the quasi-immobile ion background, energy is accumulated in the plasma and released later. If this occurs in a proper phase with the driving field and under ultra-relativistic conditions ($a_0\sim30$), a strong enhancement of the electric field is observed leading to the generation of a giant attosecond pulse [24, 28]. At our experimental conditions ($a_0\approx3$) this enhancement is less pronounced but still improves the degree of isolation of the strongest XUV pulse. Relatively low intensity is a probable reason why our experimental results do not fit the predictions of the CSE model on the harmonics spectral scaling. One of the most important differences between the models is the exponent $n$ in the spectral intensity decay power law ($I_\text{XUV}\propto\omega^{-n}$). The ROM model [10] predicts $n=8/3$ whereas in the CSE model [24] it is $n=4/3$. A fit to the experimental data presented in the paper yields an exponent of $n=2.8\pm0.3$ which fits the predictions of the ROM model well.
	
	In summary, although the frequency scaling law of the measured harmonic spectra fits the predictions of the ROM model well, the detected CEP dependence of the generated harmonics and the demonstrated possibility to generate an isolated attosecond pulse using a 3-cycle driving field are clear indications for a CSE-like behavior.% that, in principle, makes the development of a new theoretical model which includes both ROM and CSE-like processes desirable.

	%Thus in the case of a nonzero plasma scale length about $0.2\lambda-0.3\lambda$ and optimized CEP the plasma dynamics connected with the formation of electron nano-bunches acts as a gating mechanism and supports the generation of isolated attosecond pulses even with 2--3 cycle IR pulse.
	
	\section{Technical details on the streaking setup and simulations}
	
	\begin{figure}[]
		\includegraphics[width=0.7\linewidth]{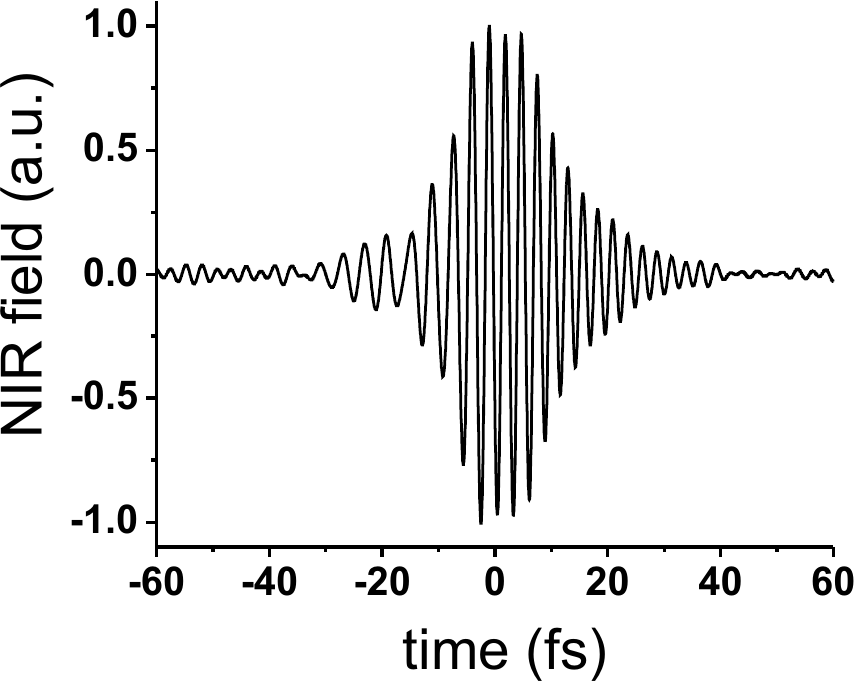}
		\caption{NIR field used in the streaking simulations.}
		\label{fig:NIR}
	\end{figure}
	
	\begin{figure}[]
		\includegraphics[width=0.9\linewidth]{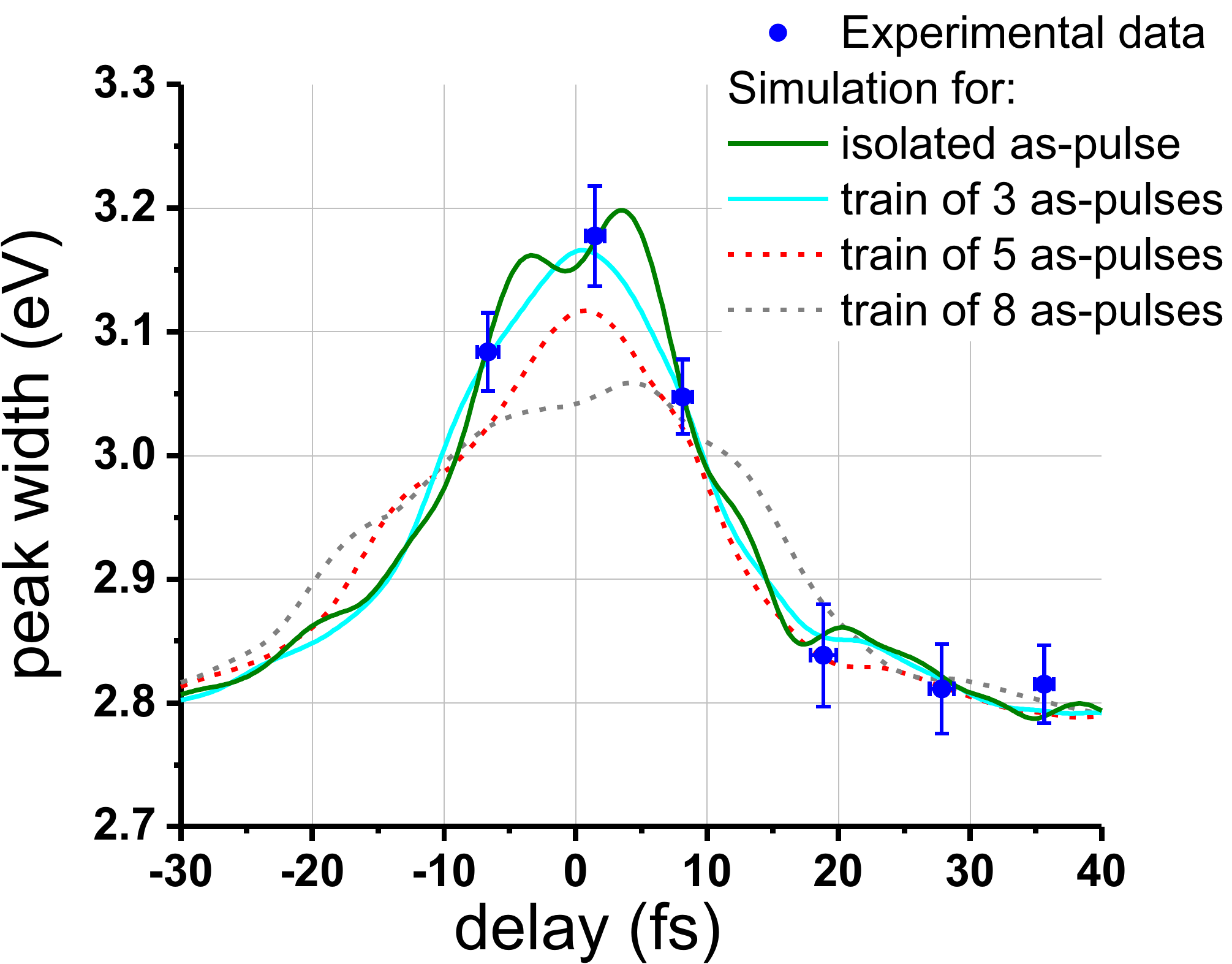}
		\caption{Results of streaking simulation assuming attosecond train with equal pulses separated by 3.3\,fs and using the NIR field presented in Fig.~\ref{fig:NIR}.}
		\label{fig:streaking}
	\end{figure}
	
	\begin{figure}[]
		\includegraphics[width=0.9\linewidth]{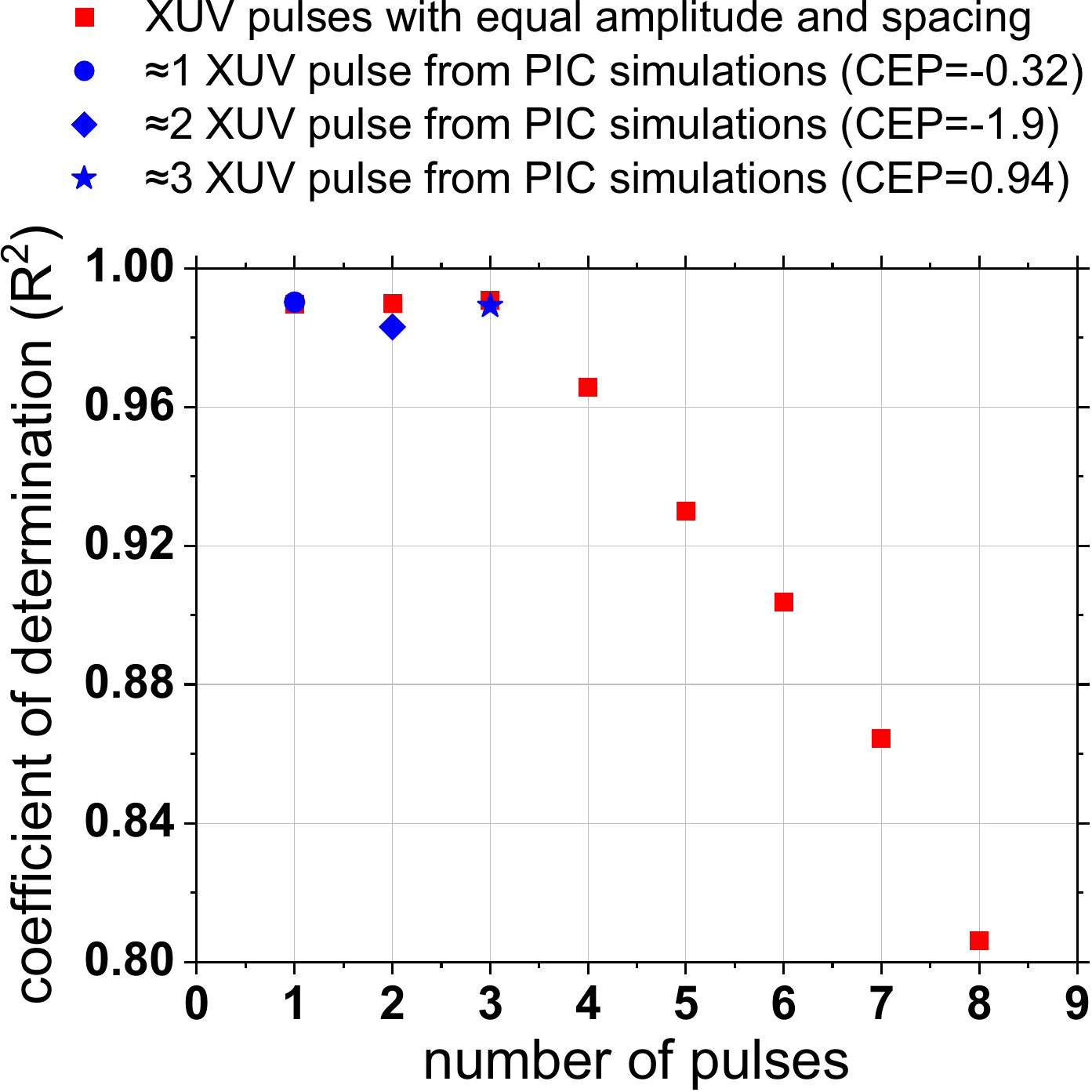}
		\caption{Goodness-of-fit parameter (coefficient of determination ($R^2$)) between experimental data and simulated streak traces for the temporal structure obtained from PIC simulation (blue dots) and for traces consisting of pulses with equal amplitude and temporal spacing (red square dots). Note that the result for another XUV pulse train model with a Gaussian envelope is similar to the presented one (red point) when the envelope of the train has the FWHM duration of $N\times T$ where N is the number of pulses and T is the pulse separation.}
		\label{fig:fit}
	\end{figure}

	In our streaking setup which is schematically presented in Fig.~1 in the paper, the beam reflected from the target is recollimated and sent into an assembly consisting of a split mirror arrangement, a neon gas jet, and a time-of-flight (TOF) photoelectron (PE) spectrometer. To record a streaking trace, the NIR and XUV pulses have to be first separated and then overlapped in space and time inside the gas jet. This is achieved with a split-mirror that consists of a stationary annular outer mirror and a 8-mm-diameter inner mirror on a piezo-electric delay stage. The outer mirror has a standard silver coating to reflect the NIR beam while the inner mirror is a specially designed multilayer XUV bandpass reflector with 3\,eV bandwidth centered at 38\,eV (XUV38BW3 from Ultrafast Innovations GmbH). Both mirrors have a focal length of $f$=25\,cm and are confocally aligned with piezo-electric actuators. In front of the split-mirror two filters where mounted: First, a pellicle with an aluminum foil in the center to block any NIR radiation incident onto the XUV mirror. The second filter is an RG780 glass with a hole in the center to block any second harmonic radiation, containing about 30\% of the total energy, incident onto the silver mirror. Passing the pellicle (2\,$\mu$m thickness) and the RG780 filter (0.5\,mm thickness) resulted in stretching of the NIR pulse to a FWHM duration of 14.5\,fs. Note that this NIR pulse elongation results in a broadening of the streaking trace but does not affect the number of generated attosecond pulses as it takes place after the SHHG process. The temporal structure of the NIR field is shown in Fig.~\ref{fig:NIR} and is used in the streaking simulations. It was calculated using the measured spectrum after reflection from the target and the dispersion of the filter provided by the supplier. The streaking traces were simulated in the strong field approximation [32] assuming a laser intensity of $10^{11}\thinspace \text{W/cm}^2$. The results for different CEP values of the NIR field were averaged in order to be able to compare the simulation with the experimental results obtained under random CEP conditions. The simulations presented in the paper were performed using the attosecond-pulse trains obtained from the PIC simulations with laser and plasma parameters corresponding to the experimental conditions. Here we additionally present (Fig.~\ref{fig:streaking}) the comparison with a simple model of a XUV train consisting of pulses with equal intensity and constant temporal spacing of 3.3\,fs. 
	
	In order to identify the number of pulses that corresponds to the most consistent streaking trace with our experimental data we used the coefficient of determination ($R^2$) as the "goodness-of-fit" parameter. $R^2$ is defined as follows:
	
	\begin{math}
	R^2=1-\frac{\sum_n(y_n-f_n)^2}{\sum_n(y_n-\overline{y})^2},
	\end{math}
	
	where y is the experimental data points; f corresponds to fit results and $\overline{y}$ is the mean value.
	
	The fit accuracy of the simulated streaking traces to the experimental data for the both the PIC-simulation case and the model-pulse-train case is shown in Fig.~\ref{fig:fit}.	The best fit accuracy of about 0.99 is provided by the XUV trains with less than 4 pulses. The simulations for more than 3 pulses in the train, for example 4 or 6, significantly deviate from the measured data and result in a worse fit accuracy of 0.96 and 0.9 correspondingly which is outside the few percent uncertainty of the $R^2$ estimation determined by the RMS error bars of the experimental data. Thus, we can conclude that the attosecond pulse train in our experiment most probably consists of not more than 3 pulses, resulting in an overall XUV train duration of $<7\thinspace \text{fs}$.

	%\section{Details on the streaking simulations}

	Compared to common streaking measurements with gas harmonics and kHz-repetition-rate driving lasers we were facing several challenges: I) lower repetition rate of 10\thinspace Hz; II) limited total number of continuous shots due to the limited target size; III) further reduction of the amount of useful data due to the necessity to select good shots (with identical performance of the laser system as described in the first section). However, the above issues can be solved in future. Namely, the points I and III can be solved with more stable and high repetition rate sources based on the thin-disc technology, while other quasi-unlimited target types such as spooling tapes should be able solve point II. During the experimental campaign, however, the limited number of shots prevented us from successfully recording a full streaking trace within the available beam time. Nevertheless the measured data are sufficient to determine an upper limit for the XUV pulse duration as described in the analysis above.